\PassOptionsToPackage{dvipsnames}{xcolor} 
\documentclass[document]{elsarticle}
\usepackage[utf8]{inputenc}
\usepackage{hyperref}
\usepackage{graphicx}
\usepackage{epstopdf, epsfig}
\usepackage{commath}
\usepackage{amsmath}
\usepackage{amssymb}
\usepackage{relsize}

\usepackage{float}
\usepackage{caption}
\usepackage{subcaption}
\usepackage{mathtools}
\usepackage{multicol}
\usepackage{dblfloatfix}
\usepackage{array}
\usepackage[title]{appendix}
\usepackage{hyperref}
\usepackage{float}
\usepackage{enumitem}

\usepackage{pdflscape}
\usepackage{relsize}

\usepackage{etoolbox}
\usepackage{tikz}
\usetikzlibrary{tikzmark}
\usetikzlibrary{calc}

\usepackage{pdflscape}
\usepackage{relsize}
\usepackage{tikz}
\usepackage{rotating}                   
\usetikzlibrary{shapes}
\usetikzlibrary{intersections}
\usetikzlibrary{positioning}
\usetikzlibrary{arrows}
\input{arrowsnew}
\usetikzlibrary{arrows.meta}
\usetikzlibrary{matrix}

\usepackage{colortbl}

\usepackage{tkz-euclide}
\usetkzobj{all}

\usepackage{makecell}
\usepackage{algorithm}
\usepackage{algpseudocode}

\algnewcommand{\LineComment}[1]{ \Statex \( \hspace{0.6cm} \triangleright\) #1}

\newcommand{\ALGtikzmarkcolor}{black}
\newcommand{\ALGtikzmarkextraindent}{4pt}
\newcommand{\ALGtikzmarkverticaloffsetstart}{-.5ex}
\newcommand{\ALGtikzmarkverticaloffsetend}{-.5ex}
\makeatletter
\newcounter{ALG@tikzmark@tempcnta}

\newcommand\ALG@tikzmark@start{%
    \global\let\ALG@tikzmark@last\ALG@tikzmark@starttext%
    \expandafter\edef\csname ALG@tikzmark@\theALG@nested\endcsname{\theALG@tikzmark@tempcnta}%
    \tikzmark{ALG@tikzmark@start@\csname ALG@tikzmark@\theALG@nested\endcsname}%
    \addtocounter{ALG@tikzmark@tempcnta}{1}%
}

\def\ALG@tikzmark@starttext{start}
\newcommand\ALG@tikzmark@end{%
    \ifx\ALG@tikzmark@last\ALG@tikzmark@starttext
    \else
        \tikzmark{ALG@tikzmark@end@\csname ALG@tikzmark@\theALG@nested\endcsname}%
        \tikz[overlay,remember picture] \draw[\ALGtikzmarkcolor] let \p{S}=($(pic cs:ALG@tikzmark@start@\csname ALG@tikzmark@\theALG@nested\endcsname)+(\ALGtikzmarkextraindent,\ALGtikzmarkverticaloffsetstart)$), \p{E}=($(pic cs:ALG@tikzmark@end@\csname ALG@tikzmark@\theALG@nested\endcsname)+(\ALGtikzmarkextraindent,\ALGtikzmarkverticaloffsetend)$) in (\x{S},\y{S})--(\x{S},\y{E});%
    \fi
    \gdef\ALG@tikzmark@last{end}%
}

\apptocmd{\ALG@beginblock}{\ALG@tikzmark@start}{}{\errmessage{failed to patch}}
\pretocmd{\ALG@endblock}{\ALG@tikzmark@end}{}{\errmessage{failed to patch}}
\makeatother

\DeclarePairedDelimiter{\ceil}{\lceil}{\rceil}
\newcolumntype{P}[1]{>{\centering\arraybackslash}p{#1}}
\newcommand{\txt}[1]{\mathsmaller{\rm{#1}}}
\newcommand{\sign}[1]{\text{sign}\left( #1\right)}
\newcommand{\defeq}{\vcentcolon=}
\newcommand{\eqdef}{=\vcentcolon}
\newcommand{\diag}[1]{\,\text{diag}\left( #1 \right)}
\newcommand{\rank}[1]{\rm{rank}\left( #1 \right)}
\newcommand{\SPAN}[1]{\rm{span} ( #1 )}
\newcommand{\bm}[1]{\boldsymbol{#1}}
\renewcommand{\min}[1]{\text{min}\left( #1 \right)}
\newcommand{\Rey}{Re}

\newtagform{Alph}[]()
\newtagform{roman}[]()
\newtagform{scroman}[][]

\newtheorem{remark}{Remark}

\newtheorem{assumption}{Assumption}

\newcommand{\bemph}[1]{{\upshape#1}} 
\newcommand{\ep}[1]{\bemph{(}#1\bemph{)}} 

\makeatletter
\DeclareRobustCommand*\uell{\mathpalette\@uell\relax}
\newcommand*\@uell[2]{
  \setbox0=\hbox{$#1\ell$}
  \setbox1=\hbox{\rotatebox{10}{$#1\ell$}}
  \dimen0=\wd0 \advance\dimen0 by -\wd1 \divide\dimen0 by 2
  \mathord{\lower 0.1ex \hbox{\kern\dimen0\unhbox1\kern\dimen0}}
}



\makeatletter
\newcommand{\algcolor}[2]{%
  \hskip-\ALG@thistlm\colorbox{#1}{\parbox{\dimexpr\linewidth-2\fboxsep}{\hskip\ALG@thistlm\relax #2}}%
}

\makeatother

\makeatletter
\def\@author#1{\g@addto@macro\elsauthors{\normalsize%
    \def\baselinestretch{1}%
    \upshape\authorsep#1\unskip\textsuperscript{%
      \ifx\@fnmark\@empty\else\unskip\sep\@fnmark\let\sep=,\fi
      \ifx\@corref\@empty\else\unskip\sep\@corref\let\sep=,\fi
      }%
    \def\authorsep{\unskip,\space}%
    \global\let\@fnmark\@empty
    \global\let\@corref\@empty  
    \global\let\sep\@empty}%
    \@eadauthor={#1}
}
\makeatother

\journal{Journal of Applied Mathematical Modelling}
\begin{document}
\begin{frontmatter}

\title{Pipe Roughness Identification of Water Distribution Networks: The Full Turbulent Case}

\author{Stefan Kaltenbacher\corref{cor1}\fnref{EMAIL}}
\author{Martin Steinberger}
\author{Martin Horn}
\address{Institute of Automation and Control\\ Graz University of Technology}
\cortext[cor1]{Address all correspondence to this author.}
\fntext[EMAIL]{\texttt{Email address:} \href{mailto:s.kaltenbacher@tugraz.at}{s.kaltenbacher@tugraz.at} }

\address{Inffeldgasse 21b, 8010 Graz, Austria}

\begin{abstract}

This paper proposes a technique to identify individual pipe roughness parameters in a water distribution network by means of the inversion of the steady-state hydraulic network equations. By enabling the reconstruction of these hydraulic friction parameters to be reliable, this technique improves the conventional model's accuracy and thereby promises to enhance model-based leakage detection and localization. 
As it is the case in so-called \textit{fireflow} tests, this methodology is founded on the premise to measure the pressure distributed at a subset of nodes in the network's graph while assuming the nodal consumption to be known. 
Beside of the proposed problem formulation, which is restricted to only allow turbulent flow in each of the network's pipes initially, developed algorithms are presented and evaluated using simulation examples.

\end{abstract}

\begin{keyword}
Roughness Calibration; Water Distribution Networks; Parameter Identification; Colebrook-White; Darcy-Weisbach; Hydraulic Friction Parameters
\end{keyword}

\end{frontmatter}

\section{Introduction}

This work is motivated by the need for efficiency improvements in the distribution of water via large hydraulic distribution networks, specifically in the drinking water supply. According to the \textit{International Water Association} \citep{IWA}, it is estimated that between a staggering 25 and 50 percent of the total water amount supplied through hydraulic networks is lost in the distribution. 
This stated percentage refers to the so-called non-revenue water which also accounts for water which is stolen, although one can expect this stolen amount to be rather insignificant.
%

Effectively, this paper proposes a new problem formulation for the model-based determination of individual roughness values per pipe in the network as well as first approaches on how to \textit{uniquely} solve it. This work is primarily focused, apart from the problem formulation, on the deduction of the concrete circumstances which allow a unique solution to this problem. 

Historically, the main research focus shifted from heuristic methods in the rather early appearance of calibration algorithms, as in \citep{cite:BhaveCalibration, cite:WalskiCalibration1983}, over explicit methods, e.g. \citep{ExplicitCalibration_Boulos, ExplicitCalibration_Ormsbee}, to implicit methods based on optimization problems, minimizing the error between measured and simulated quantities (early references are, for instance, \cite{cite:Ormsbee_ImplicitCalibration, cite:Lansey_Calibration}). Explicit methods characterize those which require to directly solve steady-state hydraulic network equations \citep{cite:Todini} for the determination of friction parameters. 
In this paper, we highlight the differences to existing approaches in literature directly at the appropriate parts. However, a more thorough literature overview can also be  found in, e.g., \citep{cite:QuoVadisCalibration}, \citep{cite:AdvancedWDModeling} or \citep{cite:Kapelan_phdthesis}.

\paragraph{Notation}
Generally, vectors and matrices are highlighted bold and italic and are consistently assigned to variables featuring lower- and upper-case letters respectively.
%
 %
%
\noindent Bold $\bm{1}_{x}$ and  $\bm{0}_{x}$ with size $x$ characterize a matrix or vector filled with ones or zeros, whereas size $x$ is only provided if it is unclear from the context. For instance, $\bm{1}_3 = [\begin{matrix} 1 &1 &1\end{matrix}]^T$ or
\[
\bm{0}_{2\times3} = 
\left[
\begin{matrix}
0 &0&0\\
0&0&0
\end{matrix}
\right] .
\]
%
\noindent The bracket-operator $[\bm{A}]_{ij} = A_{ij}$ applied on matrix $\bm{A} \in \mathbb{K}^{n\times m}$ of a number field $\mathbb{K}$, e.g. $\mathbb{K}=\mathbb{R}$ or $\mathbb{K}=\mathbb{C}$, selects element $A_{ij}$ of matrix $\bm{A}$ in row  $i\in \{1,2,\ldots,n\}$ and column $j \in \{1,2,\ldots,m\}$.
%
\noindent \sloppy Bold letter $\bm{e}_i$ utilizing index $i \in \mathbb{N}$ characterizes a unity vector $\bm{e}_i = [\begin{matrix}0 &\ldots &0 &1 &0 &\ldots &0 \end{matrix}]^T$ with variable size where $[\bm{e}_i]_{j} = 0 \,\, \forall i\ne j$ but $[\bm{e}_i]_i=1$.
\noindent The subscript  $\{-1,0,1\}$ in the set of integers $\mathbb{Z}_{\{-1,0,1\}}$ highlights that only subset $\{-1,0,1\}$ is used instead of all integers.
The ceil operator $\ceil{.}$ is applied to denote the rounding to the next higher integer. The equality symbol supplemented with double dots, as in $\text{ex}_1 \eqdef \text{ex}_2$ for instance, denotes an explicit definition which assigns the expression at the equality symbol, i.e.  $\text{ex}_1$, to the expression at the double dots, i.e. $\text{ex}_2$.


\section{Preliminaries}

\subsection{Steady-State Hydraulic Network Equations}
For this analysis the steady-state hydraulic network equations are considered which allow (among other things) to neglect unsteady friction components (see e.g. \citep{cite:Chaudhry,cite:UnsteadyFriction_Stoianov}) needed to model fast transient effects. Mathematically, the network is represented by a graph applying a set of \textit{Kirchhoff} equations, i.e. nodal equations
\begin{subequations}
\label{eq:ssEQ}
\begin{equation} \label{eq:ssEQ_1}
\bm{A} \bm{x}_Q = \bar{\bm{q}}
\end{equation}
\sloppy including the pipe flow (volumetric flow rate in m$^3$/s) vector $\bm{x}_Q = [\begin{matrix} Q_1 &Q_1 &\ldots &Q_{n_{\uell}}\end{matrix}]^T \in  \mathbb{R}^{n_{\uell}}$ of $\mathfrak{P} = \{1,2,\ldots, n_{\uell}\}$ pipes/edges,  the nodal consumption vector $\bar{\bm{q}} = [\begin{matrix} q_1 & q_2 \ldots & q_{n_{\rm{j}}}\end{matrix}]^T \in \mathbb{R}_{\ge 0}^{n_{\rm{j}}}$ of $\mathfrak{I} = \{1,2,\ldots, n_{\rm{j}} \}$ \textit{inner} nodes/vertices of the network, also considering the incidence matrix $\bm{A} \in \mathbb{Z}_{\{-1,0,1\}}^{n_{\rm{j}} \times n_{\uell}}$ which comprises minus ones, zeros and ones only. Concerning $\bm{A}$, flows influent to \textit{inner} nodes $\mathfrak{I}$ are counted positively whereas flows effluent of nodes are counted negatively. These nodal equations provide $n_{\rm{j}}$ equations out of a minimum of $n_{\uell}$ to obtain a unique flow vector $\bm{x}_Q$. The second set of $n_{\uell} - n_{\rm{j}}$ cycle equations reads as
\begin{equation} \label{eq:ssEQ_2}
\bm{S} \bm{h}_{\txt{loss}}(\bm{x}_Q) \equiv \bm{S} \tilde{\bm{C}}_s \bm{h}_s
\end{equation}
\end{subequations}
and include cycle matrix $\bm{S} \in \mathbb{Z}_{\{-1,0,1\}}^{(n_{\uell} - n_{\rm{j}}) \times n_{\uell}}$, function $\bm{h}_{\txt{loss}} : \mathbb{R}^{n_{\uell}} \rightarrow \mathbb{R}^{n_{\uell}}$ expressing hydraulic friction as pressure head losses whereas $[\bm{h}_{\txt{loss}}(\bm{x}_Q)]_i = h_{\txt{loss},i}(Q_i)$ with $i \in \mathfrak{P}$, the source (pressure) heads $\bm{h}_s \in \mathbb{R}^{n_{\rm{s}}}_{\ge 0}$ (also known as fixed heads) of $\mathfrak{S} = \{n_{\rm{j}}+1,\ldots,n_{\rm{j}}+n_{\rm{s}}\}$ source nodes and the source incidence matrix $\tilde{\bm{C}}_s \in \mathbb{Z}^{n_{\uell} \times n_{\rm{s}}}_{\{-1,0,1\}}$ (see e.g. \citep{cite:Bhave}). Generally, cycle equations \eqref{eq:ssEQ_2} satisfy \textit{Bernoulli}'s principle, also called the principle of the \textit{conservation of energy}, which says that there must be no difference in energy between two points in the network regardless of the path taken to connect these points. In other words, the sum of all head-losses along each of the network's cycles must equal zero.

 Note that according to this formulation, which has yet to be completed, $n_{\uell} \ge n_{\rm{j}}$ holds. In this context, head losses $\Delta h_i = h_{\txt{loss},i}(Q_i) = \frac{\Delta p_i}{\rho g}$ (in m) among $i \in \mathfrak{P}$, also considering the water density $\rho$ and the gravitational acceleration $g\approx 9.81$ m/s$^2$, are equivalent to the water height necessary to produce a pressure loss of $\Delta p_i$ (in Pa). Contrary to $\bm{A}$, flows effluent of \textit{source} nodes $\mathfrak{S}$ are counted positively whereas flows influent to source nodes are counted negatively concerning $\tilde{\bm{C}}_s$. This notation was kept in favor of consistency among publications, see e.g. \citep{cite:ModelingHydraulicNetworks,cite:DynamicModel_CCWI,transitionalWaterFlow}. The intersecting set of inner and source nodes is empty $\mathfrak{I} \cap \mathfrak{S} = \{ \}$ whereas their combination $\mathfrak{I} \cup \mathfrak{S} = \mathfrak{N}=\{1,2,\ldots,n_{\rm{j}}+n_{\rm{s}}\}$ yields the complete set  $\mathfrak{N}$ of the network's nodes.

\begin{assumption}[Graph]  \label{ass:Graph}
The graph representing the hydraulic network is connected and does not contain self-loops, i.e. there is no edge/pipe where starting and end node are identical. Also, the network has at least one source node $n_{\rm{s}} \ge 1$.
\end{assumption}
\begin{remark} \label{remark:AS_ortho}
Let Assumption \ref{ass:Graph} hold, then $\rank{\bm{A}} = n_{\rm{j}}$ and $\bm{S}\bm{A}^T = \bm{0}$, i.e. incidence matrix $\bm{A}$ and cycle matrix $\bm{S}$ are orthogonal. \ep{Proofs can be found e.g. in \citep{cite:GraphsBook}.}
\end{remark}
In order for the solution of \eqref{eq:ssEQ} to result in a unique flow vector $\bm{x}_Q$, friction function $\bm{h}_{\txt{loss}}(\bm{x}_Q)$ must satisfy the following properties.
\begin{assumption}[Friction Function]  \label{ass:hloss}
Friction function $\bm{h}_{\txt{loss}}(.)$ is strictly monotonically increasing, continuous and at least once continuously differentiable. It further satisfies $\bm{h}_{\mathsmaller{\rm{loss}}}(\bm{0}) = \bm{0}$.
\end{assumption}
\begin{remark}
Let Assumption \ref{ass:hloss} hold. Then, \eqref{eq:ssEQ} has a unique solution $\bm{x}_Q$ for a specific configuration with nodal consumption $\bar{\bm{q}}$ and source head $\bm{h}_s$ \cite{cite:PilatiTodini_uniqueness}. Actually, the requirement for function $\bm{h}_{\txt{loss}}(.)$ to be monotonically increasing suffices for the solution of \eqref{eq:ssEQ} to be unique \cite{cite:PillerPhD}.
\end{remark}
Knowing that $\bm{S}\bm{A}^T = \bm{0}$ according to Remark \ref{remark:AS_ortho}, i.e. $\ker(\bm{S}) = \SPAN{\bm{A}^T}$, there exists an $\bm{\alpha} \in \mathbb{R}^{n_{\rm{j}}}$ such that \eqref{eq:ssEQ_2} yields $\tilde{\bm{C}}_s \bm{h}_s - \bm{h}_{\txt{loss}}(\bm{x}_Q) = \bm{A}^T \bm{\alpha}$. \cite{cite:Todini} show that this variable $\bm{\alpha} = \bm{h} + \bm{z}$, i.e. it is equivalent to the nodal pressure head $\bm{h} \in \mathbb{R}_{\ge 0}^{n_{\rm{j}}}$ plus the geographical elevation $\bm{z} \in \mathbb{R}^{n_{\rm{j}}}_{\ge 0}$ (with respect to a common datum) of $\mathfrak{I}$ inner nodes of the network. As a remark, it is assumed that the geographical elevation of source nodes is already accommodated in the source head $\bm{h}_s$ which is contrary to the notation applied for nodal pressure heads $\bm{h}$. In sum, the solution of expressions
\begin{subequations}
\label{eq:ssTodini}
\begin{gather}
\bm{A} \bm{x}_Q = \bar{\bm{q}}\\
\bm{A}^T (\bm{h} + \bm{z}) = \tilde{\bm{C}}_s \bm{h}_s - \bm{h}_{\txt{loss}}(\bm{x}_Q) \label{eq:ssTodini_b}
\end{gather}
\end{subequations}
as it was proposed, e.g. by \cite{cite:Todini}, also provides nodal pressure head vector $\bm{h}$ in addition to flow vector $\bm{x}_Q$ when compared to \eqref{eq:ssEQ}, although \eqref{eq:ssEQ} already suffices to obtain a unique $\bm{x}_Q$.

%
\paragraph{Two-Cycle Network Example} \label{example:2Loop}
For illustrative purposes consider figure \ref{fig:2Loop},
\tikzset{
  on each segment/.style={
    decorate,
    decoration={
      show path construction,
      moveto code={},
      lineto code={
        \path [#1]
        (\tikzinputsegmentfirst) -- (\tikzinputsegmentlast);
      },
      curveto code={
        \path [#1] (\tikzinputsegmentfirst)
        .. controls
        (\tikzinputsegmentsupporta) and (\tikzinputsegmentsupportb)
        ..
        (\tikzinputsegmentlast);
      },
      closepath code={
        \path [#1]
        (\tikzinputsegmentfirst) -- (\tikzinputsegmentlast);
      },
    },
  },
  mid arrow/.style={postaction={decorate,decoration={
        markings,
        mark=at position .5 with {\arrow[#1]{stealth}}
      }}},
}

\tikzstyle{bigblock} = [draw, rectangle, minimum height=5.5cm, minimum width=10cm,rounded corners,>=latex']
   
\begin{figure}[H]
\centering
\begin{tikzpicture}[box/.style={draw,rounded corners,text width=3cm,align=center},scale=1.2]
\node[] (N1) {};
\node[] at ([yshift=-2cm]N1) (N2) {};
\node[] at ([xshift=2cm,yshift=-2cm]N1) (N3){};
\node[] at ([xshift=2cm]N1) (N4){};
\node[] at ([xshift=2cm]N4) (N5){};
\node[] at ([xshift=2cm,yshift=-2cm]N4) (N6){};
\node[] at ([xshift=-1.5cm,yshift=-1cm]N1) (R) {};
\node[] at ([xshift=-0.5cm]R) (R_LD) {};
\node[] at ([xshift=-0.5cm,yshift=0.75cm]R) (R_LU) {};
\node[] at ([xshift=0.25cm,yshift=0cm]R) (R_RD) {};
\node[] at ([xshift=0.25cm,yshift=0.75cm]R) (R_RU) {};
\node[] at ([yshift=-0.3cm]R_LU) (R_dummy) {};
\draw[thick] (R_LU) |- (R_RD);
\draw[thick] (R_RU)|- (R_LD);
\draw[fill=blue,opacity=0.3] (R_LD) rectangle ([yshift=-0.3cm]R_RU);
\draw[draw,<->,>=latex'] ($(R_LD) + (0.2,0)$) -- node[right] {\scalebox{0.85}{$h_s$}} ($(R_LD)+(0.2,0.45)$);
\node[] at ([yshift=0.2cm]N1) {\scriptsize{$k$=1}};
\node[] at ([yshift=-0.2cm,xshift=0.1cm]N3) {\scriptsize{$k$=3}};
\node[] at ([yshift=0.2cm,xshift=-0.1cm]N4) {\scriptsize{$k$=2}};
\node[] at ([xshift=-0.1cm,yshift=-0.2cm]R) {\scriptsize{Reservoir (R)}};

\node[circle,fill=black,inner sep=0pt,minimum size=5pt] (C1) at (N1) {};
\node[circle,fill=black,inner sep=0pt,minimum size=5pt] (C3) at (N3) {};
\node[circle,fill=black,inner sep=0pt,minimum size=5pt] (C4) at (N4) {};

\path [draw=black,line width= 0.025cm,postaction={on each segment={mid arrow=blue}}] (C1) -- node[above] {{\color{blue!85}\scriptsize{$Q_1$}}}(C4); 
\path [draw=black,line width= 0.025cm,postaction={on each segment={mid arrow=blue}}] (C1) |- node[] {}(C3); 
\node[] at ([xshift=-0.2cm,yshift=1cm]N2) {{\color{blue!85}\scriptsize{$Q_2$}}};

\path [draw=black,line width= 0.025cm,postaction={on each segment={mid arrow=blue}}] (C3) -- node[left] {{\color{blue!85}\scriptsize{$Q_4$}}}(C4); 
\path [draw=black,line width= 0.025cm,postaction={on each segment={mid arrow=blue}}] (C4) -|  node[above] {}($(N6)$); 
\path [draw=black,line width= 0.025cm,postaction={on each segment={mid arrow=blue}}] ($(N6)$) --  node[above] {{\color{blue!85}\scriptsize{$Q_3$}}}(C3); 
\path [draw=black,line width= 0.025cm,postaction={on each segment={mid arrow=blue}}] ($(R_RD)$) --  node[above] {{\color{blue!85}\scriptsize{$Q_5$}}}(C1); 

\draw[draw,->,>=latex'] ($(N4)$) --  ($(N4)+(0.7,0.5)$);
\node[above] at ($(N4)+(0.9,0.2)$) {\scalebox{0.85}{$q_{\txt{2}}$}};
\draw[draw,->,>=latex'] ($(N3)$) --  ($(N3)+(-0.7,-0.5)$);
\node[above] at ($(N3)+(-0.9,-0.65)$) {\scalebox{0.85}{$q_{\txt{3}}$}};

\end{tikzpicture}
\caption{Two-Cycle/Loop Network.}
\label{fig:2Loop}
\end{figure}
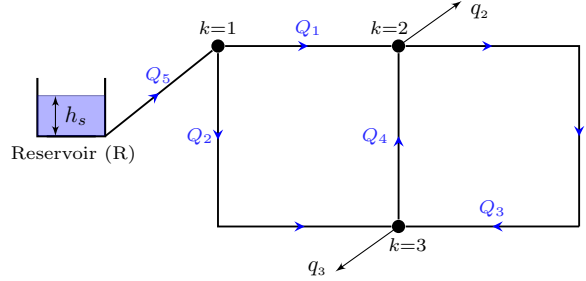
\noindent a network with $n_{\uell}=5$ pipes numbered by $\mathfrak{P}=\{1,2,\ldots,5\}$, $n_{{\rm{j}}} = 3$ \textit{inner} nodes numbered by $k\in\mathfrak{I}=\{1,2,3\}$ and one source, i.e. $n_{{\rm{s}}}=1$, providing constant pressure head $h_s$ from reservoir R. Suppose that consumers are sitting at nodes $k=2$ and $k=3$ and thereby consume $q_2$ and $q_3$ (m$^3$/s). Then,
\vspace{-0.15cm}
\begin{equation}
\underbrace{\left[ 
\begin{matrix}
-1 &-1&0&0&1\\
1&0&-1&1&0\\
0&1&1&-1&0
\end{matrix}
\right]}_{\bm{A}} \underbrace{\left[
\begin{matrix}
Q_1\\
\vdots\\
Q_5\\
\end{matrix}
\right]}_{\bm{x}_Q} =  \underbrace{\left[
\begin{matrix}
0\\
q_2\\
q_3
\end{matrix}
\right]}_{\bar{\bm{q}}},
\qquad \qquad
\tilde{\bm{C}}_s =
\left[ 
\begin{matrix}
0 &0 &0 &0 &1
\end{matrix}
\right]^T.
\label{eq:2Loop_A}
\end{equation}
The head loss over a pipe must equal the difference in the nodal pressure heads when also considering the nodal elevation. 
For the network in figure \ref{fig:2Loop} this means
\begin{equation}
\bm{h}_{\txt{loss}}(\bm{x}_Q) =\left[\begin{matrix} -\bm{A}^T & \tilde{\bm{C}}_s \end{matrix}\right]
 \left[ \begin{matrix}
\bm{h}+\bm{z}\\
\bm{h}_s
\end{matrix}\right] =
 \left[ 
\begin{matrix}
1 &1&0&0&-1\\
-1&0&1&-1&0\\
0&-1&-1&1&0\\
0&0&0&0&1
\end{matrix}
\right]^T \left[
\begin{matrix}
h_1 + z_1\\
h_2 + z_2\\
h_3 + z_3\\
h_s
\end{matrix}
\right]
\label{eq:hloss_energy}
\end{equation}
where each entry in rows of \eqref{eq:hloss_energy} characterizes the difference in nodal heads. One possibility for the cycle matrix $\bm{S}$ is
\begin{equation}
\bm{S}  = 
\left[
\begin{matrix}
1 &-1 &1 &0 &0\\
-1 & 1 &0 &1 &0
\end{matrix}
\right]
\label{eq:example_S1}
\end{equation}
which can be verified by analyzing the network in figure \ref{fig:2Loop}. This cycle matrix is generally not unique, however, one can show the network does contain $n_{\uell}-n_{\rm{j}}$ \textit{linear independent} cycles. Multiplying \eqref{eq:hloss_energy} (which is equivalent to \eqref{eq:ssTodini_b}) with the cycle matrix $\bm{S}$ from the left, one obtains  \eqref{eq:ssEQ_2} as $\bm{S}\bm{A}^T=\bm{0}$ in reference to Remark \ref{remark:AS_ortho}. Regarding term $ \bm{S} \tilde{\bm{C}}_s \bm{h}_s$ in  \eqref{eq:ssEQ_2}, cycle matrix $\bm{S}$ also accounts for linearly independent paths from one source to another such that the sum of head losses along those paths must equal the differences in source pressure heads $\bm{h}_s$, a consequence of source nodes $\mathfrak{S}$ being excluded from $\bm{A}$.
%


\subsection{Hydraulic Friction}

The pressure head loss in steady-state over a pipe due to friction along the pipe's surface and viscosity effects can be described by the \textit{Darcy-Weisbach} equation 
\begin{equation}
\Delta h_{\txt{DW}} = \lambda \frac{l \abs{Q} Q}{2d g A^2} = \lambda k \abs{Q} Q
\label{eq:DW}
\end{equation}
using the pipe's length $l$, its cross section area $A$, its diameter $d$, the gravitational acceleration $g$, and the friction factor $\lambda$ which itself again depends on pipe (volumetric) flow (rate) $Q$ and the pipe's roughness (height) $\epsilon$ usually specified in \textit{millimeters}. 
The friction factor $\lambda$ has to be distinguished between the three flow regimes, depending on the \textit{Reynolds} number
\begin{align}
\Rey = \frac{\rho d}{A \eta} \abs{Q}
\label{eq:Re}
\end{align}
also considering the water density $\rho$, and the dynamic water viscosity $\eta$. Figures in this paper are adapted to $\eta = 1.0526 \times 10^{-3}$ Pa$\cdot$s and  $\rho=998.5986$ kg/m$^3$ for a temperature of 18$^{\circ}$C.  Since the boundaries for the different flow regimes vary in literature, we stick to the ones used by \cite{cite:FluidMechanics_White,cite:Bhave,cite:AdvancedWDModeling} (referring to the \textit{Moody} diagram) specifying laminar flow below $\Rey=2000$ and turbulent flow above $\Rey=4000$.

Actually, in this research area there is lively discussion on whether to use the \textit{Hazen-Williams} equation (see e.g. \citep{cite:AdvancedWDModeling}) or \eqref{eq:DW}, although both sides agree that the description of \textit{Darcy-Weisbach} \eqref{eq:DW} in combination with the friction factor according to \textit{Colebrook} \& \textit{White} \citep{cite:ColebrookWhite} 
\begin{gather}
F_{cw}(\lambda) = \frac{1}{\sqrt{\lambda}} + \frac{2}{\ln(10)} \,\ln\left(\frac{\epsilon}{3.7 d} + \frac{2.51}{\Rey \sqrt{\lambda}} \right) = 0 \qquad \text{for} \qquad Re \ge 4000
\label{eq:CW}
\end{gather}
is more accurate and physically related. Relation \eqref{eq:CW} is actually only valid for the turbulent regime, i.e. for $\Rey \ge 4000$, and describes the friction factor $\lambda$ as the positive real solution $\lambda$ of the implicit equation $F_{cw}(\lambda)=0$ in the turbulent region.
A paper by Walski and Ormsbee with the title ``\textit{No Calm in West Palm}'' \citep{cite:NoCalmInWestPalm} specifically addresses this debate. The error made by \textit{Hazen-Williams} in comparison to \textit{Darcy-Weisbach} is allegedly so minor that its simplicity outweighs its inaccuracy. In this context, \textit{Hazen-Williams} is only valid on a narrow range of $Re$ values \citep{cite:DarcyWeisbach}. Nonetheless, the implicit \textit{Colebrook \& White} function $F_{cw}(\lambda)$ is semi-empirical and can actually be related to the partial differential \textit{Continuity} and \textit{Momentum} equations \cite{cite:FluidMechanics_White}. It is extensively considered in literature and widely established in the field of fluid mechanics and will thus be applied for further analysis.

In addition to losses due to friction along the pipe's surface and viscosity effects, so-called minor losses per pipe 
\begin{equation}
\Delta h_{m} = k_m \abs{Q} Q
\end{equation}
with an additional friction parameter $k_m$ are considered. These minor losses can be attributed to a specific point in the network rather than the losses along an entire pipe ($\Delta h_{\txt{DW}}$ is directly proportional to the pipe's length $l$) and are caused by appurtenances, such as fittings, valves (fully opened), 90 degree bends etc., penetrating the pipe and thereby provoke turbulences.
%
\begin{assumption}[Minor Losses]  \label{ass:minorLossesCalibration}
Minor losses of the $n_{\uell}$ pipes concerning $\mathfrak{P}$ in the network can be neglected.
\end{assumption}
%

With the aim to identify friction parameters, one would need to determine two parameters for each pipe in the network, namely $n_{\uell}$ roughnesses $\epsilon$ and $n_{\uell}$ minor loss parameters $k_m$. To keep the number of unknowns in a range that allows them to be uniquely reconstructed from measurement data, Assumption \ref{ass:minorLossesCalibration} is vital and very common (although often presumed implicitly) in this research field. As a consequence of Assumption \ref{ass:minorLossesCalibration}, one obtains
\begin{equation}
[\bm{h}_{\txt{loss}}(\bm{x}_Q)]_i = h_{\txt{loss},i}(Q_i) = \Delta h_i \stackrel{\eqref{eq:DW}}{=} \Delta h_{\txt{DW},i} \qquad \forall i \in \mathfrak{P}
\end{equation}
in the context of the hydraulic network equations \eqref{eq:ssTodini} and \eqref{eq:ssEQ} respectively.

\subsection{Colebrook \& White's Flow}

Although there is no explicit expression for the turbulent head loss when using $\Delta h = \Delta h_{\txt{DW}}$ according to \eqref{eq:DW} in combination with $F_{cw}(\lambda) = 0$ according to \eqref{eq:CW}, it is possible to explicitly specify the turbulent flow. Reformulating (\ref{eq:DW}) in terms of $\abs{Q}= \sqrt{\frac{\abs{\Delta h}}{\lambda k}}$ and then equating it with (\ref{eq:Re}) results in
\vspace{-0.25cm}
\begin{equation}
\Rey \sqrt{\lambda} = \frac{\rho d}{\eta A} \sqrt{\frac{\abs{\Delta h}}{k}} 
\quad  \stackrel{\eqref{eq:CW}}{\Rightarrow} \quad 
 \frac{1}{\sqrt{\lambda}} = -\frac{2}{\ln(10)} \ln\left(\frac{\epsilon}{3.7 d} + 2.51  \frac{\eta A}{\rho d} \sqrt{\frac{k}{\abs{\Delta h}}} \right)
\label{eq:1lambda} 
\vspace{-0.35cm}
\end{equation}
which can be inserted into $\abs{Q}= \frac{1}{\sqrt{\lambda}} \sqrt{\frac{\abs{\Delta h}}{k}}$, leading to turbulent \textit{Colebrook \& White}'s flow
\begin{equation}
f_t(\epsilon,\Delta h) = -\sign{\Delta h}\frac{2}{\ln(10)} \sqrt{\frac{\abs{\Delta h}}{k}} \ln\left(\frac{\epsilon}{3.7 d} + 2.51  \frac{\eta A}{\rho d} \sqrt{\frac{k}{\abs{\Delta h}}} \right).
\label{eq:ft}
\end{equation}
considered as a function on the roughness $\epsilon$ and the pressure head loss along the pipe $\Delta h$.
%
\begin{remark}
The smoothness of function \eqref{eq:ft} is compromised at $\Delta h=0$ and $\forall \epsilon$ as $\Delta h$ appears in the denominator. However, $\Delta h = 0$ already implies that pipe flow $Q$ is zero, referring to \eqref{eq:DW}. This subsequently leads to $\Rey=0$ and, hence, to laminar flow conditions per definition \ep{it is actually motionless to be precise}.
\end{remark}
%
Apart from $\Delta h = 0$, function $f_t(\epsilon,\Delta h)$ is smooth even in the laminar as well as transitional regime.

\section{Network and Sensor Configuration}

This section briefly summarizes the sensor configuration as well as further necessary assumptions in order for the roughness identification scheme to be feasible. 
 
\begin{assumption}[Known Quantities] \label{ass:known}
The pipes' dimensions, i.e. their length and diameter as well as the network's topology and the nodal elevation are known. Also, the source pressure $\bm{h}_s$ and the nodal consumption $\bar{\bm{q}}$ is assumed to be known.
\end{assumption}

Up to now, assuming the consumption to be perfectly known is, admittedly, unrealistic. However, there is some hope that customers will be equipped with direct measurement devices \citep{cite:waterConsumption} in the future. In this context, there is little doubt that wireless data transmission is inevitable, driving down maintenance- and installation costs by economies of scale. New network protocols such as LORA \citep{cite:LORA} which only demand low energy while being able to transmit data over larger distances might be able to accelerate the transition to an integrated monitoring system.

Also, in order to handle a large number of unknowns by means of $n_{{\rm{p}}} < n_{{\rm{j}}}$ pressure sensors measuring
\begin{equation}
\label{eq:Ch}
\bm{y}_h = \bm{C}_h \bm{h}, \qquad \text{with} \qquad
\bm{C}_h = \left[
\begin{matrix}
 \bm{e}_{p_1} & \bm{e}_{p_2} &\ldots &\bm{e}_{p_{n_{{\rm{p}}}}} 
\end{matrix}
\right]^T
\in \mathbb{Z}^{n_{{\rm{p}}} \times n_{{\rm{j}}}}_{\{0,1\}}
\end{equation}
at the subset $\mathcal{P}=\{p_1,p_2,\ldots,p_{n_{\rm{p}}}\} \subseteq \mathfrak{I}$, several sets of measurements, which have to be taken during different loading conditions (concerning $\bm{h}_s$ and $\bar{\bm{q}}$), are needed.
\begin{assumption}[Linear Independency] \label{ass:linearIndependence}
There are $n_{{\rm{m}}}$ sets of sufficiently \textit{linear independent} measurements, denoted by $\mathfrak{M} = \{1,2,\ldots,n_{\rm{m}}\}$, available. Linear independence can be achieved by a variation of source pressure $\bm{h}_s$ and/or the nodal consumption $\bar{\bm{q}}$.
\end{assumption}

Since consumers are currently not equipped with the necessary sensor technology, so-called \textit{fireflow} tests are usually conducted. At the minimum-night-flow, somewhere from 2am to 5am where the regular water consumption is lowest, hydrants are opened systematically. The hydrants' outflow is measured as well as pressure sensors distributed throughout the network record, at least part of, the pressure distribution. Depending on the amount of the minimum-night-flow, this procedure is, potentially, very problematic for calibration as nodal \textit{Kirchhoff} equations (conservation of mass) are violated when the sum of all considered hydrant flows (fireflows) is lower than the total inflow in the network.

\begin{assumption}[Steady-State] \label{ass:SteadyStateCalibration}
These $n_{{\rm{m}}}$ sets of measurements are taken in \textit{steady-state} of the network.
\end{assumption}

In order to avoid distortions due to transient effects, which have not been considered in the modeling procedure, the network has to be in steady-state during the time-frame considered for each of the measurement-sets. Recording a larger time-frame with a number of different measurement values for each sensor in each measurement-set may also be valuable for filtering noise. Applying simple averaging often proves effective in this regard.

\begin{assumption}[Noise] \label{ass:NoiseCalibration}
The variance of the measurement noise of applied pressure sensors is significantly smaller than the pressure drop, i.e. $ [{\text{var}}(\bm{y}_h^{(i)})]_k \ll [\tilde{\bm{C}}_s \bm{h}_s^{(i)} - \bm{A}^T (\bm{h}^{(i)} + \bm{z})]_j$ for all $k =1,2,\ldots,n_{\rm{p}}$ and $j \in \mathfrak{I}$ in at least one of the $i \in \mathfrak{M}$ measurement-sets. This means that the inequality holds for all $j,k$ in at least one of the $i$-th measurement-sets. Also, the measurement errors in fireflows and pressures have zero mean.
\end{assumption}

One can sum up all assumptions in table \ref{tab:calibrationAssumptions}.
\begin{table}[H]
\begin{center}
\begin{tabular} { P{70pt} | P{220pt}  } 
		Assumption                        & Context                   \\ \hline\hline
		\ref{ass:Graph}                &  properties of the graph   	  	      \\ \hline 
		\ref{ass:hloss}                  &  characteristics of $\bm{h}_{\txt{loss}}(.)$ 		\\ \hline
		\ref{ass:minorLossesCalibration}   & negligible minor losses\\  \hline
	        \ref{ass:known}              & pipe dimensions, source pressure, consumption	\\ \hline
\ref{ass:linearIndependence}        & independency of measurements\\  \hline
\ref{ass:SteadyStateCalibration}   & measurements in steady-state\\  \hline
\ref{ass:NoiseCalibration}              & measurement noise \\  \hline
\end{tabular}
\caption{Summary of assumptions relevant for roughness calibration.}
\label{tab:calibrationAssumptions}
\end{center}
\end{table}
%
%
%

\section{Full Turbulent Problem Set-Up}
Following \eqref{eq:Ch}, $n_{\rm{p}}$ out of $n_{\rm{j}}$ nodal pressure heads are measured which means that $n_{\rm{j}} - n_{\rm{p}}$ not-measured pressure heads
\begin{equation} \label{eq:Chb}
\bm{h}_N = \bar{\bm{C}}_h \bm{h} \qquad \text{with} \qquad \bar{\bm{C}}_h = 
\left[
\begin{matrix}
 \bm{e}_{\bar{p}_1} & \bm{e}_{\bar{p}_2} & \ldots & \bm{e}_{\bar{p}_{n_{{\rm{j}}}-n_{{\rm{p}}}}}
\end{matrix}
\right]^T
\end{equation}
have to be considered unknown. Indices of unity vectors $\bm{e}_i \in \mathbb{Z}^{n_j}_{\{0,1\}}$ thereby belong to subset $i \in \bar{\mathcal{P}}=\{\bar{p}_1,\bar{p}_2,\ldots,\bar{p}_{n_{{\rm{j}}}-n_{{\rm{p}}}}\}\subseteq \mathfrak{I}$ such that $\mathcal{P} \cap \bar{\mathcal{P}} = \{\}$ and $\mathcal{P} \cup \bar{\mathcal{P}} = \mathfrak{I}$. 

%
\begin{assumption} \label{ass:FullTurbulent}
Suppose that each pipe flow $j$ in each measurement-set $i$  is in the turbulent regime, i.e.
\begin{equation}
\label{eq:Re_bla}
Re^{(i)}_j =\frac{|Q^{(i)}_j|  d_j \rho}{A_j \eta} \ge 4000 \qquad \forall j \in \mathfrak{P} \land \forall i \in \mathfrak{M}
\end{equation}
\end{assumption}
Assumption \ref{ass:FullTurbulent} is necessary for the Colebrook-White flow \eqref{eq:ft} to be generally applicable. 

\subsection{Problem Statement}
Combining the network equations \eqref{eq:ssTodini} with \textit{Colebrook \& White}'s flow \eqref{eq:ft} while considering \eqref{eq:Ch} and \eqref{eq:Chb}, the nodal equations along the $i$-th measurement-set have the following structure
%
\begin{subequations}
\label{eq:calibrationTur}
\begin{gather}
\bm{A} \bm{x}_Q(\bm{\epsilon}, \bm{h}_N^{(i)}) = \bar{\bm{q}}^{(i)} \label{eq:calibrationTur_1}\\
\Delta\bm{h}^{(i)} = \tilde{\bm{C}}_s \bm{h}_s^{(i)} - \bm{A}^T \bm{C}^T_h \bm{y}_h^{(i)} - \bm{A}^T \bar{\bm{C}}_h^T \bm{h}_N^{(i)} - \bm{A}^T\bm{z} \label{eq:calibrationTur_2}\\
[\bm{x}_Q(\bm{\epsilon}, \Delta\bm{h}^{(i)})]_j =Q^{(i)}_j \stackrel{\eqref{eq:ft}}{=} f_{t,j}([\bm{\epsilon}]_j, [\Delta\bm{h}^{(i)}]_j) \quad  \forall j \in \mathfrak{P} 
\label{eq:calibrationTur_3}
\end{gather}
\end{subequations}
for all $i\in \{1,2,\ldots, n_{{\rm{m}}}\} =\mathfrak{M}$ where the $j$-th flow component \eqref{eq:calibrationTur_3} used for \textit{Kirchhoff} equations \eqref{eq:calibrationTur_1} is calculated via \eqref{eq:ft}, i.e. the flow in the turbulent regime, thereby applying the conservation of energy for the head losses \eqref{eq:calibrationTur_2}.
 Effectively, the unknowns of this set of equations are the roughnesses $\bm{\epsilon} \in \mathbb{R}^{n_{\uell}}_{\ge 0}$ and the pressure heads at nodes with no sensors $\bm{h}_N^{(i)} \in \mathbb{R}^{n_{{\rm{j}}}-n_{{\rm{p}}}}_{\ge 0}$ in the $i$-th measurement-set. 

At this point \eqref{eq:calibrationTur} contains, in principle, all information needed to determine all the pipes' roughness, provided that the assumptions in table \ref{tab:calibrationAssumptions} hold and sufficiently many measurement-sets are available. Emphasizing on the application of the \textit{Colebrook} \& \textit{White} formula with respect to the explicit turbulent flow expression, \eqref{eq:ft} has not been considered for any explicit or implicit calibration scheme in literature as far as the authors are aware. A more recent proposal for an explicit approach applying \textit{Hazen-William}'s description of pipe friction can be found in, for instance, \citep{cite:explicitCal_1}. However, \eqref{eq:calibrationTur} turns out to be particularly difficult to solve even in the unperturbed case when no measurement noise is considered. The reason for that can not only be attributed to the problem's size, which is considerably larger than the steady-state network equations \eqref{eq:ssTodini}, but to the nonlinear dependency of \eqref{eq:ft} on $\bm{h}_N^{(i)}$.

\subsection{Number of Measurement-Sets}

The principle of the presented approach starts with the premise to obtain at least as many equations as unknowns. In the first set of measurements, that is $i=1$, there are only $n_{\rm{j}}$ nodal \textit{Kirchhoff} equations to accommodate $n_{\uell}+n_{\rm{j}}-n_{\rm{p}}$ unknowns. Important to note here is that the independent cycle equations (conservation of energy) \eqref{eq:calibrationTur_2} have to be used implicitly for the set-up of nodal equations and thus provide no additional information.

The key observation is that the nodal \textit{Kirchhoff} equations in the second set of measurements, i.e. $i=2$, are independent of the nodal equations from the first measurement-set given Assumption \ref{ass:linearIndependence}. However, as in the second measurement-set only $n_{\rm{j}}-n_{\rm{p}}$ \textit{additional} unknown pressure heads have to be considered, the \textit{additional} $n_{\rm{j}}$ independent nodal equations improve the relation between the combined number of equations $2n_j$ to the number of unknowns $n_{\uell} + 2 (n_{\rm{j}} - n_{\rm{p}})$. The number of nodal equations grows faster than the number of unknown pressure heads $n_{{\rm{j}}} > n_{{\rm{j}}} - n_{{\rm{p}}}$ with each measurement-set.
 
Given a fixed number of sensors $n_{{\rm{p}}}$, the minimum number of measurement-sets in order to reach a break-even point is denoted with $n_{{\rm{m,min}}}$. One obtains
\begin{gather}
[n_{{\rm{m,min}}} (n_{{\rm{j}}}-n_{{\rm{p}}}) + n_{\uell}] - n_{{\rm{m,min}}} n_{{\rm{j}}} < 0 \nonumber\\
n_{{\rm{m,min}}} = \ceil*{n_{\uell}/n_{{\rm{p}}}}  ,
\end{gather}
meaning that $n_{\uell}/n_{\rm{p}}$ has to be rounded to the next higher integer. Interestingly, this requirement is completely independent of the number of nodes, junctions (more than two connections) to be precise, in the network. Nonetheless, one should not be deceived by this result since the number of equations grows linearly with $n_{{\rm{j}}}$, making the solving more difficult due to numerical issues.

\section{A Modified \textit{Newton-Raphson} Approach} \label{sec:Newton}

With the aim to solve a nonlinear set of equations of the form $\bm{f}(\bm{x}) \stackrel{!}{=} \bm{0}$
with a smooth and continuous vector function $\bm{f}: \mathbb{R}^{n} \rightarrow \mathbb{R}^{n}$, an iterative scheme $\bm{x}_k = \bm{x}_{k-1}+\Delta \bm{x}_k$ along iterations denoted by $k$ provides \textit{quadratic} convergence to the real root $\bm{x}^* \Rightarrow \bm{f}(\bm{x}^*) = \bm{0}$ if the initial value $\bm{x}_0$ is chosen in vicinity of $\bm{x}^*$. This also provides that the full search direction
\begingroup
\allowdisplaybreaks
\begin{equation}
\Delta \bm{x}_k =\bm{x}_{k}-\bm{x}_{k-1}= -\left(  \left. \frac{\partial \bm{f}}{\partial \bm{x}} \right|_{\bm{x}_{k-1}} \right)^{-1} \bm{f}(\bm{x}_{k-1}) = - \bm{J}_{k-1}^{-1} \bm{f}(\bm{x}_{k-1})
\label{eq:NewtonDirection}
\end{equation}
with step length $\mu=1$ is taken in all iterations, requiring the inverse of \textit{Jacobian} $\bm{J}_{k-1}$ to be square and have full rank. However, in the present application \eqref{eq:calibrationTur}
\begin{subequations}
\label{eq:fx_xQTur_1}
\begin{align}
\bm{f}(\bm{x}) &= \left[
\begin{matrix}
\bm{A} & &\\
 &  \ddots&\\
& &  \bm{A}
\end{matrix}
\right] \left[
\begin{matrix}
\bm{x}_Q(\bm{\epsilon}, \bm{h}_N^{(1)})\\
\bm{x}_Q(\bm{\epsilon}, \bm{h}_N^{(2)})\\
\vdots\\
\bm{x}_Q(\bm{\epsilon}, \bm{h}_N^{(n_{{\rm{m}}})})
\end{matrix}
\right] - \left[
\begin{matrix}
\bar{\bm{q}}^{(1)}\\
\bar{\bm{q}}^{(2)}\\
\vdots\\
\bar{\bm{q}}^{(n_{{\rm{m}}})}\\
\end{matrix}
\right] \quad 
\text{with} \\
 \bm{x}^T &= \left[
\begin{matrix}
\bm{\epsilon}^T & \bm{h}_N^{(1)^T} & \ldots & \bm{h}_N^{(n_{{\rm{m}}})^T}
\end{matrix}
\right] 
\raisetag{10pt}
\end{align}
\end{subequations}
\endgroup
function $\bm{f} : \mathbb{R}^{n_{{\rm{m}}} n_{{\rm{j}}}} \rightarrow \mathbb{R}^{n_{\uell} + n_{{\rm{m}}} (n_{{\rm{j}}} - n_{{\rm{p}}})}$ has, in general, not the same number of components as the number of variables, i.e. $n_{{\rm{m}}} n_{{\rm{j}}} \ne n_{\uell} + n_{{\rm{m}}} (n_{{\rm{j}}} - n_{{\rm{p}}})$. Taking a closer look at its thin (more rows than columns) \textit{Jacobian}
\begin{equation}
\bm{J}(\bm{x}) =  \left[
\begin{matrix}
\bm{A} & &\\
 &  \ddots&\\
& &  \bm{A}
\end{matrix}
\right] \left[
\begin{matrix}
\frac{\partial \bm{x}_Q(\bm{\epsilon},\bm{h}_N^{(1)})}{\partial \bm{\epsilon}} & \frac{\partial \bm{x}_Q(\bm{\epsilon},\bm{h}_N^{(1)})}{\partial \bm{h}_N^{(1)}} &\bm{0} &\ldots &\bm{0} \\
\frac{\partial \bm{x}_Q(\bm{\epsilon},\bm{h}_N^{(2)})}{\partial \bm{\epsilon}} & \bm{0}&  \frac{\partial \bm{x}_Q(\bm{\epsilon},\bm{h}_N^{(2)})}{\partial \bm{h}_N^{(2)}}  &\ldots &\bm{0}\\
\vdots & \vdots & &\ddots\\
\frac{\partial \bm{x}_Q(\bm{\epsilon},\bm{h}_N^{(n_{{\rm{m}}})})}{\partial \bm{\epsilon}} & \bm{0}&  \bm{0}&  \ldots &\frac{\partial \bm{x}_Q(\bm{\epsilon},\bm{h}_N^{(n_{{\rm{m}}})})}{\partial \bm{h}_N^{(n_{{\rm{m}}})}}  
\end{matrix}
\right],
\label{eq:Jacobian}
\end{equation}
where $\bm{J} \in \mathbb{R}^{n_{{\rm{m}}} n_{{\rm{j}}} \times n_{\uell} + n_{{\rm{m}}} (n_{{\rm{j}}} - n_{{\rm{p}}})}$ is supposed to have full $\rank{\bm{J}(\bm{x})} =  n_{\uell} + n_{{\rm{m}}} (n_{{\rm{j}}} - n_{{\rm{p}}})$ in reference to Assumption \ref{ass:linearIndependence}. Note that $\bm{J}_{k-1} \defeq \bm{J}(\bm{x}_{k-1})$. A quite practical possibility to deal with the non-square form of $\bm{J}(\bm{x})$ is to take the left inverse
$\bm{J}^{+} \defeq (\bm{J}^T \bm{J})^{-1} \bm{J}^T$.
%

\subsection{First Turbulent Flow Derivatives} \label{sec:FirstDerivatives}
In order to build up the \textit{Jacobian} according to \eqref{eq:Jacobian}, the derivatives of \eqref{eq:ft} with respect to $\bm\epsilon$ and $\bm{h}_N$ are needed. To display these derivatives more compactly, the argument of the natural logarithm in \eqref{eq:ft}
 \begin{equation}
\ell = \ell(\epsilon,\Delta h) = \frac{\epsilon}{3.7 d} + 2.51 \frac{\eta A}{\rho d} \sqrt{\frac{k}{\abs{\Delta h}}}
\label{eq:ell}
\end{equation}
is denoted by $\ell$. 
Starting with the roughness, one obtains (neglecting indices on the right hand side of \eqref{eq:dQde_scalar})
\begin{subequations}
\label{eq:dQdx}
\begin{equation}
\frac{\partial f_{t,i}}{\partial \epsilon_i} \eqdef p_{\epsilon,i} (\epsilon_i, \Delta h_i) \mathrel{\hat=} \frac{\partial f_{t}}{\partial \epsilon} = -\frac{2}{\ln{(10)}} \text{sign}(\Delta h) \sqrt{\frac{\abs{\Delta h}}{k}}   \frac{1}{3.7 d  \, \ell(\epsilon, \Delta h)} \quad \forall i\in \mathfrak{P}
\label{eq:dQde_scalar}
\end{equation}
followed by $\frac{\partial f_{t,i}}{\partial h_{N,j}} = \frac{\partial f_{t,i}}{\partial \Delta h_i} \frac{\partial \Delta h_i}{\partial h_{N,j}} \eqdef p_{\Delta h, i} (\epsilon_i,\Delta h_i) \frac{\partial \Delta h_i}{\partial h_{N,j}} \quad \forall i\in \mathfrak{P} \land j \in \bar{\mathcal{P}} \mathrel{\hat=}$
\begin{equation}
\frac{\partial f_t}{\partial h_N} = - \frac{1}{\ln{(10)}} \left( \sqrt{\frac{1}{k \abs{\Delta h}}} \ln{(\ell(\epsilon,\Delta h))} - 2.51 \frac{\eta A}{\rho d} \frac{\abs{\Delta h}^{-1}}{\ell{(\epsilon, \Delta h})}\right) \frac{\partial \Delta h}{\partial h_N} 
\label{eq:dQdhN_scalar}
\end{equation}
\end{subequations}
(neglecting indices) where the partial derivative of $\Delta h$ in respect to $h_N$ is constant due to 
\begin{equation}
\label{eq:dhdhN}
\frac{\partial \Delta \bm{h}}{\partial \bm{h}_N} \stackrel{\eqref{eq:calibrationTur_2}}{=} -\bm{A}^T \bar{\bm{C}}_h, \,\,\,
\begin{aligned} 
 \quad [\Delta \bm{h}]_i &=[\bm{h}_{\txt{loss}}(\bm{x}_Q)]_i = \Delta h_i \,\,\forall i\in \mathfrak{P} \\
 \quad [\bm{h}_N]_j &= [\bar{\bm{C}}_h \bm{h}]_j = h_{N,\bar{p}_j} \,\, \forall \bar{p}_j\in \bar{\mathcal{P}} \land j \in \{1,\ldots,n_{\rm{j}}-n_{\rm{p}}\}
\end{aligned}
 \end{equation}
when considering vector dependencies. As a remark, note that the authors assumed that $\frac{\partial}{\partial \Delta h} \text{sign}(\Delta h) = 0$ neglecting the \textit{Dirac-Delta} $\delta(\Delta h)$ function. The scalar partial derivatives \eqref{eq:dQdx} can now be summarized in vector notation as follows
\begin{equation}
\label{eq:dxQ_der}
\begin{aligned}
\frac{\partial \bm{x}_Q(\bm{\epsilon},\Delta \bm{h}^{(i)})}{\partial \bm{\epsilon}} &= \diag{\bm{p}_{\epsilon}(\bm{\epsilon},\Delta \bm{h}^{(i)})} \\
 \frac{\partial \bm{x}_Q(\bm{\epsilon},\Delta \bm{h}^{(i)})}{\partial \bm{h}_N^{(i)}} &= -\diag{\bm{p}_{\Delta h}(\bm{\epsilon},\Delta \bm{h}^{(i)})} \bm{A}^T \bar{\bm{C}}_h^T
\end{aligned}
\quad \forall i \in \mathfrak{M}=\{1,2,\ldots n_{{\rm{m}}}\} 
\end{equation}
where $[\bm{p}_{\epsilon}]_j \stackrel{\eqref{eq:dQde_scalar}}{=} p_{\epsilon,j}(\epsilon_j,\Delta h_j)$ and $[\bm{p}_{\Delta h}]_j \stackrel{\eqref{eq:dQdhN_scalar}}{=} p_{\Delta h,j}(\epsilon_j,\Delta h_j)$ for all $j \in \mathfrak{P}$. Strictly speaking, pipe parameters in \eqref{eq:dQdx} would also require a pipe index, e.g. $d_j$, as they do certainly vary with each pipe. However, index $j$ was omitted to improve readability. Actually, one can recognize that the information concerning $\bm{J}$ which varies along the $i$-th measurement-sets can entirely be stored in vectors by means of $\bm{p}_{\epsilon}$ and $\bm{p}_{\Delta h}$.


\subsection{Step Length}

To relax the requirement to already start in the vicinity of $\bm{x}^*$ a suitable selection of the step length $\mu$ concerning $\bm{x}_k = \bm{x}_{k-1}+\mu \Delta \bm{x}_k$ in each iteration is needed. The authors implemented a methodology similar to the one proposed by \citep[section 9.7]{cite:numericalRecipesC}. The principle is described briefly in the following. 
 
The idea is to choose $\mu$ such that a norm, i.e. $\Vert \bm{f}(\bm{x}_k) \Vert$, decreases with each iteration step, i.e. $\Vert \bm{f}(\bm{x}_{k}) \Vert < \Vert \bm{f}(\bm{x}_{k-1}) \Vert$.  Although \citep[section 9.7]{cite:numericalRecipesC} proposes to use the $\mathcal{L}_2$ norm
\begingroup
\allowdisplaybreaks
\begin{subequations}
\label{eq:norms}
\begin{gather}
\Vert \bm{f}\Vert_{_{\mathcal{L}_2}}^2 = f_1^2 + f_2^2 + \ldots + f_{n_{{\rm{m}}} n_{{\rm{j}}}}^2 \label{eq:L2_norm}\\
\intertext{also $\mathcal{L}_1$ and $\mathcal{L}_{\infty}$ norms in the form}
\Vert \bm{f}\Vert_{_{\mathcal{L}_1}} = \sum_{p=1}^{n_{{\rm{m}}} n_{{\rm{j}}}} \abs{f_p} \label{eq:L1_norm}\\
\Vert \bm{f}\Vert_{_{\mathcal{L}_{\infty}}} = \underset{p}{\text{max}}\, \abs{f_p} \label{eq:Linf_norm}
\end{gather}
\end{subequations}
\endgroup
were tested, whereas the $\mathcal{L}_{1}$ norm turned out favorable. Details are discussed by means of an example  in the next section. 
This $\mathcal{L}_1$ norm (in contrast to \citep{cite:numericalRecipesC}) was then selected as quality-measure for the step length. 

\begin{remark} \label{remark:abs_eps}
The applied norm was designed to be symmetrical along the roughness axes by taking the absolute value of $\epsilon$ in function \eqref{eq:ft} concerning $\Vert \bm{f}(\bm{x})\Vert$ and \eqref{eq:calibrationTur} to facilitate convergence towards positive thus physical relevant roughnesses by preserving a context-type shape \ep{details in the example section}.
\end{remark}

Knowing that every root of $\bm{f}(\bm{x})$, i.e. $\bm{x}^*$, is a minimum of $v(\bm{x}) = \Vert \bm{f}(\bm{x})\Vert_{_{\mathcal{L}_1}}$, it is clear that the \textit{Newton} direction \eqref{eq:NewtonDirection} represents a \textit{descent} direction of $v(\bm{x})$, i.e.
 \begin{equation}
 \frac{\partial v}{\partial \bm{x}} \Delta \bm{x} = \frac{\partial v}{\partial \bm{f}} \frac{\partial \bm{f}}{\partial \bm{x}} \Delta \bm{x} = -\text{sign}(\bm{f})^T \bm{J} \bm{J}^{-1} \bm{f} = -\text{sign}(\bm{f})^T \bm{f} < 0 \quad \forall \bm{f} \ne \bm{0}.
 \label{eq:descentRate}
 \end{equation}
The strategy is comprised of three basic steps.
\begin{enumerate}[label=(\Roman*)]
  \item try the full $\mu = 1$ Newton step which will provide quadratic convergence eventually
  \item check at each iteration if the proposed step reduces the norm (or similar criteria)
  \item if not, backtrack along the Newton direction until an acceptable step is obtained
\end{enumerate} 

The goal is to find a $\mu \in ]0,1]$ for which 
\begin{equation} \label{eq:v_x_norm}
g(\mu) \defeq \Vert \bm{f}(\bm{x}_{k-1} + \mu \Delta \bm{x}_{k})\Vert_{_{\mathcal{L}_1}} = v(\bm{x}_{k-1} + \mu \Delta \bm{x}_{k})
\end{equation}
decreases sufficiently, that is the case for $\mu = 1$ if, e.g., the criterion \citep{cite:numericalRecipesC}
\begin{equation}
v(\bm{x}_k) \le v(\bm{x}_{k-1}) + 10^{-4} \times \left.\frac{\partial v}{\partial \bm{x}}\right|_{\bm{x}_{k-1}} \Delta \bm{x}_k,
\label{eq:StepLengthCriteria}
\end{equation}
is met. If not, one is looking for an interpolation of $g(\mu)$ with a polynomial of second degree using the function evaluation $g(0)$ from the previous step, $g(1)$ from the full Newton step $\mu=1$ as well as
\begin{equation}
g\prime(0)= \left. \frac{\partial g(\bm{x}_k)}{\partial \mu} \right|_{\mu=0} =  \left.\frac{\partial v}{\partial \bm{x}}\right|_{\bm{x}_{k-1}} \Delta \bm{x}_k = -\text{sign}(\bm{f}(\bm{x}_{k-1}))^T \bm{f}(\bm{x}_{k-1})
\end{equation}
to determine the polynomial coefficients. The $\mu^*$ which then minimizes this second-order polynomial is the next candidate for finding a new Newton step $\bm{x}_k=\bm{x}_{k-1} + \mu^* \Delta \bm{x}_k$ which suffices \eqref{eq:StepLengthCriteria}. However, if this $\bm{x}_k$ then again does not comply with  \eqref{eq:StepLengthCriteria}, one takes the new, additional evaluation of $g(\mu)$, that is $g(\mu^*)$, to determine the coefficients of a third-order polynomial interpolating $g(\mu)$. The minimum $\mu^+$ of this third-order polynomial is then the candidate for the next step length. Details are found in  \citep[section 9.7]{cite:numericalRecipesC} and in Algorithm \ref{alg:Newton} on page \pageref{alg:Newton}.

\begin{remark} \label{remark:ScalingNewton} 
Since the argument of equation set \eqref{eq:calibrationTur} to be solved has two sets of components, namely $\bm{\epsilon}$ and $\bm{h}_N^{(i)}$, which are \ep{at least} in the range of 3 orders of magnitude \ep{SI units} different from each other, it is advisable to scale $\bm{x}\in \mathbb{R}^{n_{\uell}+n_{{\rm{m}}} (n_{{\rm{j}}}-n_{{\rm{p}}})}$ for the Newton direction calculation. Numerical issues become dominant with growing number of nodes $n_{{\rm{j}}}$ and pipes $n_{\uell}$. 
\end{remark}

The scaling, in reference to Remark \ref{remark:ScalingNewton}, was not included in Algorithm \ref{alg:Newton} in order to keep the complexity reasonable for illustrative purposes.


\begin{algorithm}[H] 
\caption{Modified \textit{Newton-Raphson} algorithm with step length variation}  \label{alg:Newton}
\begin{algorithmic}[1]
\Procedure{Newton}{$\text{FUN},\bm{x}_0$} \Comment{FUN characterizes a pointer on a }
 \LineComment{function returning the residuum of \eqref{eq:fx_xQTur_1} under \eqref{eq:calibrationTur} and $\bm{J}$ \eqref{eq:Jacobian}}
\Statex{ }  \hspace{0.55cm}\textsc{Initial Phase} \hrulefill
   \State{$[\bm{f}_{k},\bm{J}_{k}] \gets \text{FUN}(\bm{x}_0)$} \Comment{first function call of FUN} 
   \State $v_k \gets  \sum_{p=1}^{n_{{\rm{j}}} n_{{\rm{m}}}} \abs{f_{k,p}}$ \Comment{norm calculation \eqref{eq:L1_norm}}
   \State{$v_{k-1} \gets v_k$} \Comment{for initialization purposes only}
   \State $\Delta \bm{x}_k \gets -(\bm{J}_k^T\bm{J}_k)^{-1} \bm{J}_k^T  \bm{f}_{k}$ \label{algline:NS_1} \Comment{Newton direction with the left inverse of $\bm{J}$}
   \State{$\mu \gets 1$}
   \State{$iter \gets 0$} \Comment{iter $\ldots$ number of ``\textit{Newton}'' iterations}
\Statex{ }  \hspace{0.55cm}\textsc{Main Loop}\hrulefill   
   \While{$(\abs{v_k-v_{k-1}} > \epsilon_f$  or $\Vert \mu\Delta  \bm{x}_k \Vert_{\mathcal{L}_2} > \epsilon_x$) and ($iter <$  max $iter$)} 
\Statex{ }  \hspace{1cm} \textsc{Newton Direction}\hrulefill   
        \If{$\mu = 1$}
           \State $\Delta \bm{x}_k \gets -(\bm{J}_k^T\bm{J}_k)^{-1} \bm{J}_k^T \bm{f}_{k}$ \label{algline:NS_2}\Comment{Newton direction}
           \State$s_k \gets -\text{sign}(\bm{f}_{k})^T \bm{f}_{k}$ \Comment{rate of descent  \eqref{eq:descentRate}}
           \State$v_k \gets  \sum_{p=1}^{n_{{\rm{j}}} n_{{\rm{m}}}} \abs{f_{k,p}}$ \Comment{$\mathcal{L}_1$ norm calculation}
           
           \State{$\bm{f}_{k-1} \gets \bm{f}_k$, $\bm{x}_{k-1} \gets \bm{x}_k$, $\bm{v}_{k-1} \gets \bm{v}_k$}  \Comment{buffer old values} 
           \State{$iter \gets iter+1$} \label{algline:NewtonIteration}
        \EndIf   
\Statex{ }  \hspace{1.15cm}\textsc{Newton Step}\hrulefill        
        \State $\bm{x}_k \gets \bm{x}_{k-1} + \mu \Delta \bm{x}_k$ \Comment{next Newton Step}
        \State{$[\bm{x}_k]_j = x_{k,j} \gets \abs{x_{k,j}} \quad \text{for} \,\, j=1,2,\ldots,n_{\uell}$} \Comment{see Remark \ref{remark:abs_eps}}
        \State{$[\bm{f}_{k},\bm{J}_{k}] \gets \text{FUN}(\bm{x}_k)$}
        \State $v_k \gets  \sum_{p=1}^{n_{{\rm{j}}} n_{{\rm{m}}}} \abs{f_{k,p}}$ \Comment{norm calculation  \eqref{eq:L1_norm}}
        \State{$\mu_{\txt{old}} \gets \mu$} \Comment{buffer old step length}
\Statex{ }   \hspace{1.15cm}\textsc{Step Length Control}\hrulefill               
        \If{$v_k > v_{k-1} + 10^{-4} \mu s_k$} \Comment{criterion \eqref{eq:StepLengthCriteria}}
        	  \If{$\mu = 1$}
        	  	\State{$\mu \gets \frac{-s_k}{2 (v_k  - v_{k-1} - s_k)}$} \Comment{minimum of 2nd order polynomial}
        	 \Else\\
        	 	\hspace{2.3cm}calculate new coefficients $a,b$ for cubic polynomial with\\
        	 	\hspace{2.3cm}$\mu^* = \mu_{\txt{old}},\quad g(0) = v_{k-1},\quad g(1) = v_k, \quad g\prime(0) = s_k$
        	 	\If{$a = 0$}
        	 		\State{$\mu \gets -\frac{s_k}{2b}$} \Comment{minimum of cubic if first coefficient $a=0$}
        	 	\Else
        	 		\State{$\mu \gets \frac{-b + \sqrt{b^2 - 3 a s_k}}{3 a}$} \Comment{minimum of cubic}
        	 	\EndIf
        	 	\State{$\mu \gets \min{\mu,0.5 \mu_{\txt{old}}} $}   \Comment{maximal step length}
        	 \EndIf
        	 \State{$\mu \gets \max{(\mu,0.1 \mu_{\txt{old}})} $} \Comment{minimal step length}
        \Else
        	 \State{$\mu \gets 1$}	 
        \EndIf  
        
\Statex{ }  \vspace{-0.3cm}\hspace{1.15cm}\hrulefill                            
   \EndWhile
\Statex{ }  \vspace{-0.3cm} \hspace{0.6cm}\hrulefill                               
   \State \textbf{return} $[\bm{x}_{k-1},\bm{f}_{k-1}]$
\EndProcedure
\end{algorithmic}
\end{algorithm}

%
\subsection{Initial Values and Range}
In the sensor-noise-free case, Algorithm \ref{alg:Newton} occasionally finds the real root $\bm{x}^*$ of \eqref{eq:calibrationTur} if $\bm{x}_0$ is already close to $\bm{x}^*$. The convergence strongly depends on the initial values $\bm{x}_0$ with which the algorithm is launched. Thereby, \eqref{eq:calibrationTur} turns out to be particularly sensitive with respect to the not-measured pressures $\bm{h}_N^{(i)}$. In this context it is utterly important to define a physically useful range
\begin{subequations}
\label{eq:hNrange}
\begin{gather}
h^{(i)}_{N,\bar{p}} \in [\underline{h_{N}}_{,\bar{p}}^{(i)}, \overline{h_{N}}_{,\bar{p}}^{(i)}] \quad \forall  \bar{p} \in \bar{\mathcal{P}} \land \forall i \in \mathfrak{M}\\
\Rightarrow \bm{h}_N^{(i)} \in [\underline{\bm{h}_N}^{(i)}, \overline{\bm{h}_N}^{(i)}] \quad \forall  i \in \mathfrak{M}\\
\Rightarrow \bm{x}_{h_N} = [\begin{matrix}\bm{h}_N^{{(1)}^T} &\bm{h}_{N}^{{(2)}^T}&\ldots&\bm{h}_N^{{(n_{{\rm{m}}})}^T} \end{matrix}]^T \in [\underline{\bm{h}_N}, \overline{\bm{h}_N}] \label{eq:hNrange3}
\end{gather}
\end{subequations}
and let $\bm{h}_{N}^{(i)}$ concerning $\bm{x}_0$ (the initial value) be in this range. Otherwise the solution space of $v(\bm{x}) = \Vert \bm{f}(\bm{x})  \Vert_{\mathcal{L}_1}$ will most unlikely feature a desired convex-type form (only for $[\bm{x}]_i = \epsilon_i > 0 \,\, \forall i \in \mathfrak{P}$ in reference to Remark \ref{remark:abs_eps}). In analogy, the physically useful range for the roughnesses ought to be between 0\% and 5\% of the pipe's diameter in reference to the \textit{Moody}-chart.

In order to increase the chance of converging to the real root, the strategy to launch Algorithm \ref{alg:Newton} several times with different initial values $\bm{x}_0$ turns out successful. However, going from one initial value to another, it is useful to remember the temporarily ``best'' solution, i.e. $\bm{x}^+$, meaning the one which has the smallest residual of \eqref{eq:calibrationTur} measured by $v(\bm{x}^+)$. Thereby, the $\bm{h}_N^{+,(i)}$-components of the temporarily best solution, in terms of the smallest $v(\bm{x}^+)$, are used for the $n_{{\rm{m}}}(n_{{\rm{j}}}-n_{{\rm{p}}})$ components of the next initial value, i.e.
\begin{equation}
\bm{x}_0 = \left[
\begin{matrix}
 \bm{\epsilon}^T_0 
 & \bm{h}_N^{{+,(1)}^T}
 &\ldots
 &\bm{h}_N^{{+,(n_{{\rm{m}}})}^T}
\end{matrix}
\right]^T.
\label{eq:nextIC}
\end{equation}
Note that yet another index for \eqref{eq:nextIC} to denote the iteration along different initial values was omitted. The selection of the (next) initial roughness $\bm{\epsilon}_0$ is done by a random number generator, assuming a uniformly distributed roughness between 0\% and 5\% of the corresponding pipe's diameter. In this context, it turned out effective to vary only those elements of $\bm{\epsilon}_0$ which are not in the physically relevant range, i.e. $\epsilon_{0,i} = \text{random}(0,0.05 d_i)$ for $[\bm{x}^+]_i = \epsilon_i^+ > 0.05 d_i \quad \forall i \in \mathfrak{P}$. 

In case the Algorithm \ref{alg:Newton} does return $\bm{h}^{(i)}_N$ outside its considered range \eqref{eq:hNrange}, the returned $\bm{x}$ will not be buffered in $\bm{x}^+$, even if $v(\bm{x})$ would be the smallest so far.
Actually, Algorithm \ref{alg:IC} (on page \pageref{alg:IC}) steers $\bm{x}$, provided by Algorithm \ref{alg:Newton}, back to its physical range by varying roughnesses. Although Algorithm \ref{alg:IC} requires the not-measured pressure heads to remain inside their physical relevant range, that is $\underline{\bm{h}_N} \le \bm{x}_{h_N} \le \overline{\bm{h}_N}$ (see line \ref{alg:IC_ifCondition} of Algorithm \ref{alg:IC}), roughnesses $[\bm{x}^+]_i = \epsilon_i\,\, \forall i\in \mathfrak{P}$ can, in fact, exceed the 5\% mark of the pipe's diameter $d_i$. Variants of Algorithm \ref{alg:IC} where roughnesses, concerning $\bm{x}^+$, are forced to never exceed this $0.05 d_i=0.05 [\bm{\mathfrak{d}}]_i \,\, \forall i \in \mathfrak{P}$ boundary turned out far too conservative in the solution finding.

\begin{algorithm}[H] 
\caption{Variation of initial values}  \label{alg:IC}
\begin{algorithmic}[1]
\Procedure{NetCalibration}{$\text{FUN},\bm{x}_0$} 
\Comment{argument FUN characterizes a }
 \LineComment{pointer on a function returning the residuum of \eqref{eq:fx_xQTur_1} under \eqref{eq:calibrationTur}}
\Statex{ } \hspace{0.55cm}\textsc{Initial Phase} \hrulefill
   \State{$[\bm{x}^{+},\bm{f}^{+}] \gets \textsc{Newton}(\text{FUN},\bm{x}_0)$} \Comment{call Algorithm \ref{alg:Newton}} \label{algline:IC_newtonCall}
   \State $v^+ \gets  \sum_{p=1}^{n_{{\rm{j}}} n_{{\rm{m}}}} \abs{f^+_{p}}$ \Comment{norm calculation  \eqref{eq:L1_norm}}
   \State $\bm{x} \gets \bm{x}_0$ \Comment{for initialization only} 
   \State{$iter \gets 0$} \Comment{iter $\ldots$ number of iterations}   
\Statex{ }  \hspace{0.55cm}\textsc{Main Loop}\hrulefill   
   \While{$(v^+> \epsilon_f$ or $\Vert \bm{x} - \bm{x}^+\Vert_{\mathcal{L}_2} > \epsilon_x$) and ($iter <$  max \# of $iter$)} \label{alg:IC_whileCondition}
   
	\State{$\bm{x}_0 \gets \bm{x}^+$}
	\State{determine indices $\mathfrak{P}_{d}=[\begin{matrix} i_1 &i_2 &\ldots&i_{n_{\epsilon}}\end{matrix}]$ where}
	\Statex{\hspace{0.95cm} $[\bm{x}^+]_{i\in\mathfrak{P}} =\epsilon_i > 0.05 d_i \quad \forall i \in\mathfrak{P}_{d}$ }
	\State{$[\bm{x}_0]_i \gets \text{random}(0,0.05 d_i) \quad \forall i\in\mathfrak{P}_d$} \label{algline:index_1}\Comment{random number: $[0,5]\%$ of $d_i$}   
   
   \State{$[\bm{x},\bm{f}] \gets \textsc{Newton}(\text{FUN},\bm{x}_0)$} \label{algline:index_2}
   \State $v \gets  \sum_{p=1}^{n_{{\rm{j}}} n_{{\rm{m}}}} \abs{f_{p}}$ \Comment{norm calculation  \eqref{eq:L1_norm}}
    \State{$[\bm{x}_{h_N}]_l \gets [\bm{x}]_{l+n_{\uell}} \quad \text{for} \quad l=1,2,\ldots,n_{{\rm{m}}}(n_{{\rm{j}}}-n_{{\rm{p}}})$} \Comment{cf. \eqref{eq:hNrange3}}
   
\Statex{ }  \hspace{1.15cm}\textsc{Buffer ``Good'' Solutions}\hrulefill               
        \If{$v \le v^+$ and $\underline{\bm{h}_N} \le \bm{x}_{h_N} \le \overline{\bm{h}_N}$} \label{alg:IC_ifCondition}
                    
		\State{$\bm{x}^+ \gets \bm{x}, v^+ \gets v, \bm{f}^+ \gets \bm{f}$}	 
        \EndIf  
        
\Statex{ }  \vspace{-0.3cm}\hspace{1.15cm}\hrulefill      

      \State{$iter \gets iter+1$}
   \EndWhile
\Statex{ }  \vspace{-0.3cm} \hspace{0.6cm}\hrulefill                               
   \State \textbf{return} $[\bm{x}^{+},\bm{f}^+]$
\EndProcedure
\end{algorithmic}
\end{algorithm}

However, even when considering no disturbances at all, the real root $\bm{x}^*$ will not lead to a perfect zero, i.e. $v(\bm{x}^*) > 0$ due to numerics. A basic assumption for Algorithm \ref{alg:IC} to work is that this real root $\bm{x}^*$ has a clearly distinguishable (cf. with \eqref{eq:hNrange3})
\begin{equation}
v(\bm{x}^*) < v(\bm{x}) \quad \forall \bm{x} \in 
\left[
\left[
\begin{matrix}
\bm{0}_{n_{\uell}} \\
\underline{\bm{h}_N}
\end{matrix}
\right],
\left[
\begin{matrix}
0.05 \bm{\mathfrak{d}}\\
\overline{\bm{h}_N}
\end{matrix}
\right]
\right]
\end{equation}
value in the defined range at least. This can only be the case if Assumption \ref{ass:linearIndependence} holds, providing measurement-sets which are sufficiently independent from each other. 

The limits for Algorithm \ref{alg:IC} to abort, referring to $\epsilon_f$ and $\epsilon_x$, should actually be chosen conservatively, compared to the ones used for Algorithm \ref{alg:Newton}, to avoid too many iterations in this outer loop. Thereby, condition $v^+ > \epsilon_f$ (m$^3$/s) allows direct adjustment of the accuracy with respect to the sum of all nodal consumption-errors (heavily dependent on $n_{{\rm{j}}}$).
Also, mind that at this point one still has to select initial conditions for Algorithm \ref{alg:IC} as well as the physically relevant range for the not-measured pressures $\bm{h}_N^{(i)} \, \forall i$, namely $\underline{\bm{h}_N}$ and $\overline{\bm{h}_N}$. With the purpose to clarify the general methodology, an example is provided in the following.

\section{Simulation Example}

For illustrative purposes consider  figure \ref{fig:3CycleNet}, a network with $n_{{\rm{j}}}  = 5$ nodes, $n_{\uell}=8$ pipes, hence $n_{\uell}-n_{{\rm{j}}} = 3$  independent cycles, $n_{\rm{p}}=3$ pressure sensors and the requirement of at least $n_{\rm{m,min}} = \ceil{n_{\uell}/n_{\rm{p}}} = 3$ measurement-sets.
This network also features $n_{{\rm{q}}}=3$ consumers at nodes $k=2,3,4$ and $n_{{\rm{s}}}=1$ constant pressure source. The fact that the (red colored) nodes equipped with pressure sensors also have consumers does not affect the identification scheme.

\vspace{0.2cm}
\tikzset{
  on each segment/.style={
    decorate,
    decoration={
      show path construction,
      moveto code={},
      lineto code={
        \path [#1]
        (\tikzinputsegmentfirst) -- (\tikzinputsegmentlast);
      },
      curveto code={
        \path [#1] (\tikzinputsegmentfirst)
        .. controls
        (\tikzinputsegmentsupporta) and (\tikzinputsegmentsupportb)
        ..
        (\tikzinputsegmentlast);
      },
      closepath code={
        \path [#1]
        (\tikzinputsegmentfirst) -- (\tikzinputsegmentlast);
      },
    },
  },
  mid arrow/.style={postaction={decorate,decoration={
        markings,
        mark=at position .5 with {\arrow[#1]{stealth}}
      }}},
}

\tikzstyle{bigblock} = [draw, rectangle, minimum height=5.5cm, minimum width=10cm,rounded corners,>=latex']
   
\begingroup
\setlength{\intextsep}{2pt}%
\setlength{\columnsep}{0pt}%
\begin{figure}[H]
\begin{center}
\begin{tikzpicture}[box/.style={draw,rounded corners,text width=3cm,align=center},scale=1.2]
\node[] (N1) {};
\node[] at ([yshift=-2cm]N1) (N2) {};
\node[] at ([xshift=2cm,yshift=-2cm]N1) (N3){};
\node[] at ([xshift=2cm]N1) (N4){};
\node[] at ([xshift=2cm]N4) (N5){};
\node[] at ([xshift=2cm,yshift=-2cm]N4) (N6){};
\node[] at ([yshift=-1cm]N4) (N7){};
\node[] at ([yshift=-1cm]N5) (N8){};
\node[] at ([xshift=-1.5cm,yshift=-1cm]N1) (R) {};
\node[] at ([xshift=-0.5cm]R) (R_LD) {};
\node[] at ([xshift=-0.5cm,yshift=0.75cm]R) (R_LU) {};
\node[] at ([xshift=0.25cm,yshift=0cm]R) (R_RD) {};
\node[] at ([xshift=0.25cm,yshift=0.75cm]R) (R_RU) {};
\node[] at ([yshift=-0.3cm]R_LU) (R_dummy) {};
\draw[thick] (R_LU) |- (R_RD);
\draw[thick] (R_RU)|- (R_LD);
\draw[fill=blue,opacity=0.3] (R_LD) rectangle ([yshift=-0.3cm]R_RU);
\draw[draw,<->,>=latex'] ($(R_LD) + (0.2,0)$) -- node[right] {{$h_s$}} ($(R_LD)+(0.2,0.45)$);
\node[] at ([yshift=0.2cm]N1) {\scriptsize{$k$=1}};
\node[] at ([yshift=-0.2cm,xshift=0.1cm]N3) {\scriptsize{$k$=3}};
\node[] at ([yshift=0.2cm,xshift=-0.1cm]N4) {\scriptsize{$k$=2}};
\node[] at ([xshift=0.35cm,yshift=-0.05cm]N8) {\scriptsize{$k$=4}};
\node[] at ([xshift=-0.35cm]N7) {\scriptsize{$k$=5}};
\node[] at ([xshift=-0.2cm,yshift=-0.2cm]R) {\scriptsize{Reservoir (R)}};

\node[circle,fill=black,inner sep=0pt,minimum size=5pt] (C1) at (N1) {};
\node[circle,fill=red,inner sep=0pt,minimum size=7pt] (C3) at (N3) {};
\node[circle,fill=red,inner sep=0pt,minimum size=7pt] (C4) at (N4) {};
\node[circle,fill=black,inner sep=0pt,minimum size=5pt] (C7) at (N7) {};
\node[circle,fill=red,inner sep=0pt,minimum size=7pt] (C8) at (N8) {};

\path [draw=black,line width= 0.025cm,postaction={on each segment={mid arrow=blue}}] (C1) -- node[above] {{\color{blue!85}\scriptsize{$Q_2$}}}(C4); 
\path [draw=black,line width= 0.025cm,postaction={on each segment={mid arrow=blue}}] (C1) |- node[] {}(C3); 
\node[] at ([xshift=-0.2cm,yshift=1cm]N2) {{\color{blue!85}\scriptsize{$Q_3$}}};

\path [draw=black,line width= 0.025cm,postaction={on each segment={mid arrow=blue}}] (C7) -- node[left] {{\color{blue!85}\scriptsize{$Q_8$}}}(C3); 
\path [draw=black,line width= 0.025cm,postaction={on each segment={mid arrow=blue}}] (C4) -- node[left] {{\color{blue!85}\scriptsize{$Q_5$}}}(C7);
\path [draw=black,line width= 0.025cm,postaction={on each segment={mid arrow=blue}}] (C4) -|   node[right] {{\color{blue!85}\scriptsize{$Q_4$}}}($(C8)$); 
\path [draw=black,line width= 0.025cm,postaction={on each segment={mid arrow=blue}}] ($(C8)$) |-  node[below] {{\color{blue!85}\scriptsize{$Q_7$}}}(C3); 
\path [draw=black,line width= 0.025cm,postaction={on each segment={mid arrow=blue}}] ($(C7)$) --  node[below] {{\color{blue!85}\scriptsize{$Q_6$}}}(C8); 
\path [draw=black,line width= 0.025cm,postaction={on each segment={mid arrow=blue}}] ($(R_RD)$) --  node[above] {{\color{blue!85}\scriptsize{$Q_1$}}}(C1); 

\node[circle,fill=red,inner sep=0pt,minimum size=7pt] (C8) at (N8) {};

\draw[draw,->,>=latex'] ($(N4)$) --  ($(N4)+(0.7,0.5)$);
\node[above] at ($(N4)+(0.9,0.2)$) {{$q_2$}};
\draw[draw,->,>=latex'] ($(N3)$) --  ($(N3)+(-0.7,-0.5)$);
\node[above] at ($(N3)+(-0.9,-0.65)$) {{$q_3$}};

\draw[draw,->,>=latex'] ($(N8)$) --  ($(N8)+(0.7,0.5)$);
\node[above] at ($(N8)+(0.9,0.2)$) {{$q_4$}};

\end{tikzpicture}
\caption{3-cycle network with pressure sensors at red colored nodes $k=2,3,4$.}
\label{fig:3CycleNet}
\end{center}
\end{figure}
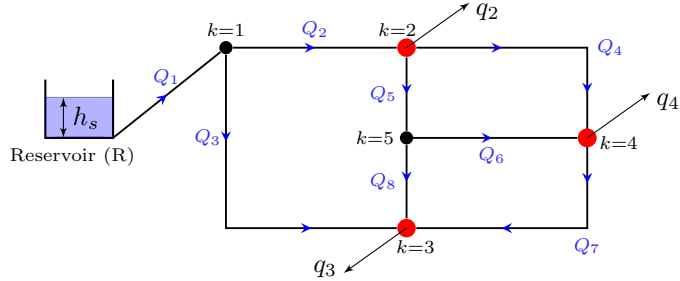
%
\vspace{-0.3cm}

First, incidence matrix
\begin{equation}
\bm{A} = \left[
\begin{matrix}
1 & -1 & -1 & 0&0&0&0&0\\
0& 1&0&-1&-1&0&0&0\\
0&0&1&0&0&0&1&1\\
0&0&0&1&0&1&-1&0\\
0&0&0&0&1&-1&0&-1
\end{matrix}
\right] \in \mathbb{Z}_{\{-1,0,1\}}^{n_{{\rm{j}}} \times n_{\uell}},
\end{equation}
nodal elevation $\bm{z} = [\begin{matrix}0 &10 &5 &0 &0 \end{matrix}]^T$ (in m), the pipes' diameter $\bm{\mathfrak{d}} = 0.04 \times \bm{1}_{n_{\uell}}$ (in m) (i.e. $d_i = [\bm{\mathfrak{d}}]_i \,\, \forall i \in \mathfrak{P}$), their length $\bm{l} = [\begin{matrix} 10 &10&20&15&5&10&15&5\end{matrix}]^T$ (in m), roughnesses $\bm{\epsilon} = [\begin{matrix}2 &1.75 &1.5&1.25&1&0.75&0.5&0.25 \end{matrix}]^T \times 10^{-3}$ (in m) are chosen, whereas minor losses are set to zero. The \textit{Colebrook} \& \textit{White} formula \eqref{eq:CW} is applied for the calculation of the friction factor $\lambda$ \eqref{eq:DW}. The solving of the implicit equation is thereby achieved iteratively. In order to produce an independent set of steady-state configurations (``measurements''), a dynamic model is utilized which has been derived in  \cite{cite:ModelingHydraulicNetworks,cite:DynamicModel_CCWI} while varying the desired consumption $\bm{q}_d$. 

For some background information concerning this dynamic model, orifice coefficients, serving as control variables for the consumption $\bm{q} \in \mathbb{R}_{> 0}^{n_{\rm{q}}}$, are not in saturation $[\bm{u}]_i = [\bm{r}]_i^{-2} \in ]1,\epsilon_r^{-2}[$ for all $i=1,\ldots,n_{\rm{q}}$. Eigenvalues are selected as $\bm{\lambda}_q = -15 \times \bm{1}_{3}$. 
However, this is not overly important for this example due to Assumption \ref{ass:SteadyStateCalibration} requiring the network to be in steady-state in each of the considered measurement-sets $\mathfrak{M}$ anyways. This means, effectively, that the dynamic equations proposed in \cite{cite:ModelingHydraulicNetworks} already converged to the equilibrium as a solution of \eqref{eq:ssTodini} in each measurement set. In this context, the equivalence of the solution of \eqref{eq:ssTodini} to the equilibrium of the dynamic model has been proven in \cite[Theorem 2]{cite:ModelingHydraulicNetworks}.

The following matrices are utilized
\begin{equation}
\bm{C}_h = \left[
\begin{matrix}
     0     &1     &0     &0    & 0\\
     0     &0    & 1     &0     &0\\
     0     &0     &0     &1    &0
\end{matrix}
\right], \qquad 
\bar{\bm{C}}_h = 
\left[
\begin{matrix}
     1     &0     &0     &0     &0\\
     0     &0    & 0     &0    & 1
\end{matrix}
\right], \qquad
\bm{C}_s = 
\left[
\begin{matrix}
1 &\bm{0}_{7}^T
\end{matrix}
\right]^T.
\end{equation}

\vspace{-0.4cm}
\subsection{Step Length}
The 3-cycle network in figure \ref{fig:3CycleNet} was chosen for analysis of the norms \eqref{eq:norms} as quality-measure for the step length. $n_{{\rm{m}}}=4$ measurement-sets were generated with varying consumption. 
In order to allow graphical representation in 3 dimensions with $n_{\uell} + n_{{\rm{m}}} (n_{{\rm{j}}}-n_{{\rm{p}}})=16$ unknowns, 14 of these unknowns were fixed in the real root $\bm{x}^*$ of \eqref{eq:calibrationTur} whereas the solution space concerning \eqref{eq:norms} along the two remaining variables was considered. Figure \ref{fig:SolSpace} allows comparison of the different norms of $\bm{f}(\bm{x})$ concerning problem \eqref{eq:calibrationTur}. The limits for the $h_{N,5}^{(3)}$ axes in figure \ref{fig:SolSpace} were determined with the help of measurements $\bm{y}_h^{(3)}$ at node $2,3,4$ such that $h_{N,5}^{(3)} \in [\underset{p}{\text{min}}\, (y_{h,p}^{(3)}),\underset{p}{\text{max}}\, (y_{h,p}^{(3)})]$ which is feasible as no sources are directly connected to node 5, see figure \ref{fig:3CycleNet}.\\

\begin{figure}[H]
 \centering
        \begin{subfigure}[H]{1\textwidth}
            \centering
            \includegraphics[width=1\linewidth]{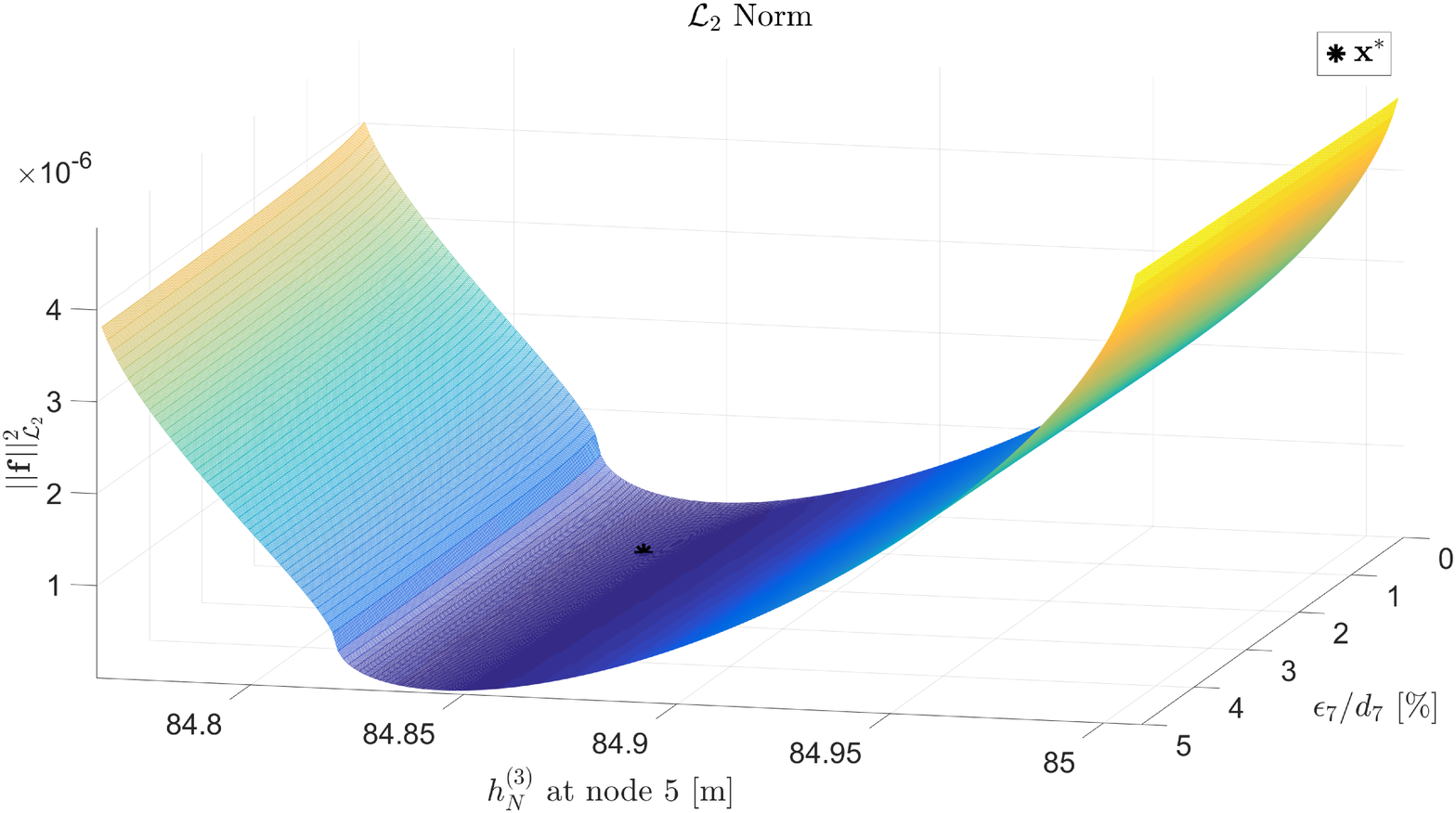}
            \caption{{\small $\mathcal{L}_2$ Norm.}}    
            \label{fig:SolSpace1}
        \end{subfigure}
\end{figure}
\begin{figure}[H]\ContinuedFloat
        \begin{subfigure}[H]{1\textwidth}  
            \centering 
            \includegraphics[width=1\linewidth]{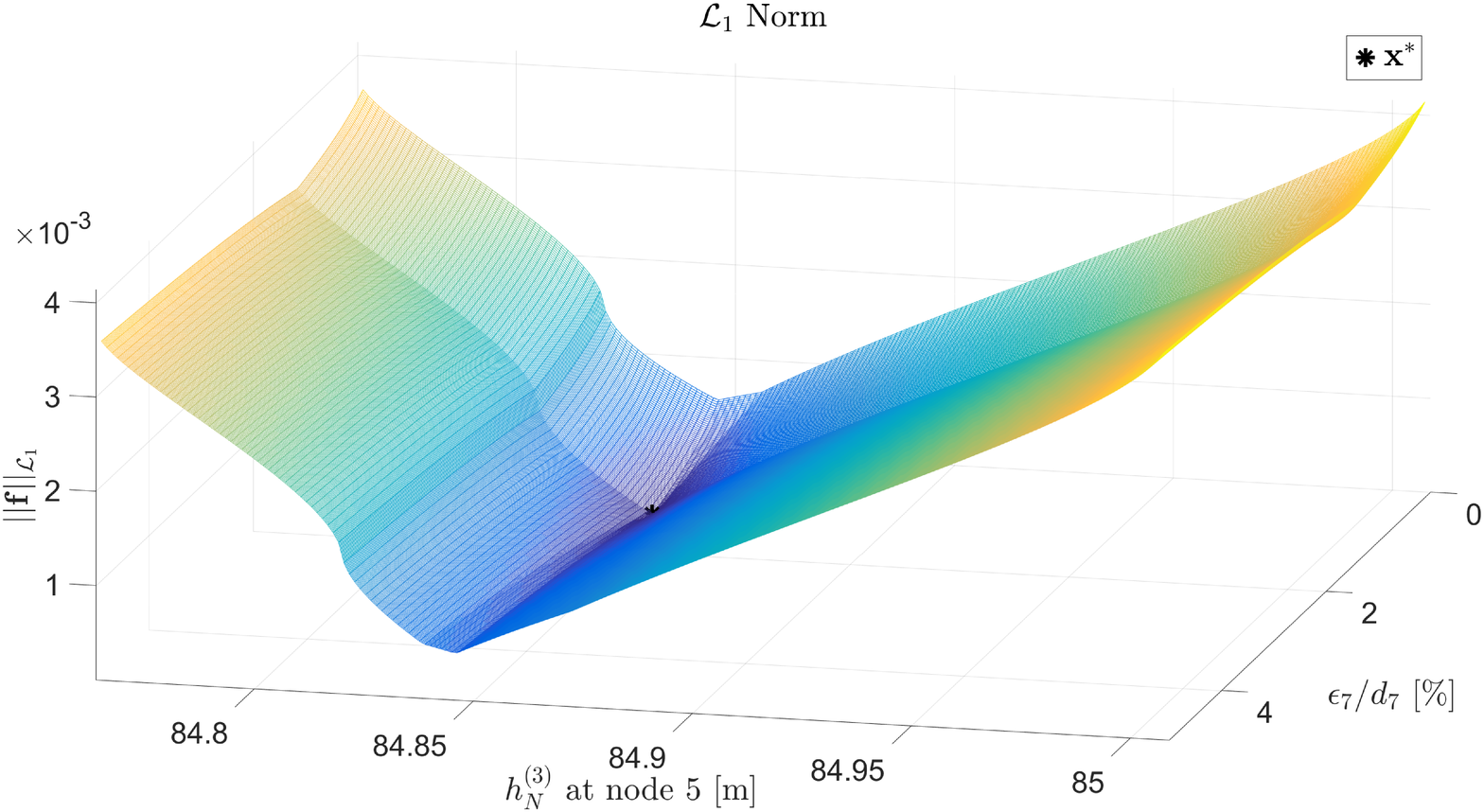}
            \caption{{\small $\mathcal{L}_1$ Norm.}}    
            \label{fig:SolSpace2}
      \end{subfigure}
\end{figure}
\begin{figure}[H]\ContinuedFloat    
        \begin{subfigure}[H]{1\textwidth}   
            \centering 
            \includegraphics[width=1\linewidth]{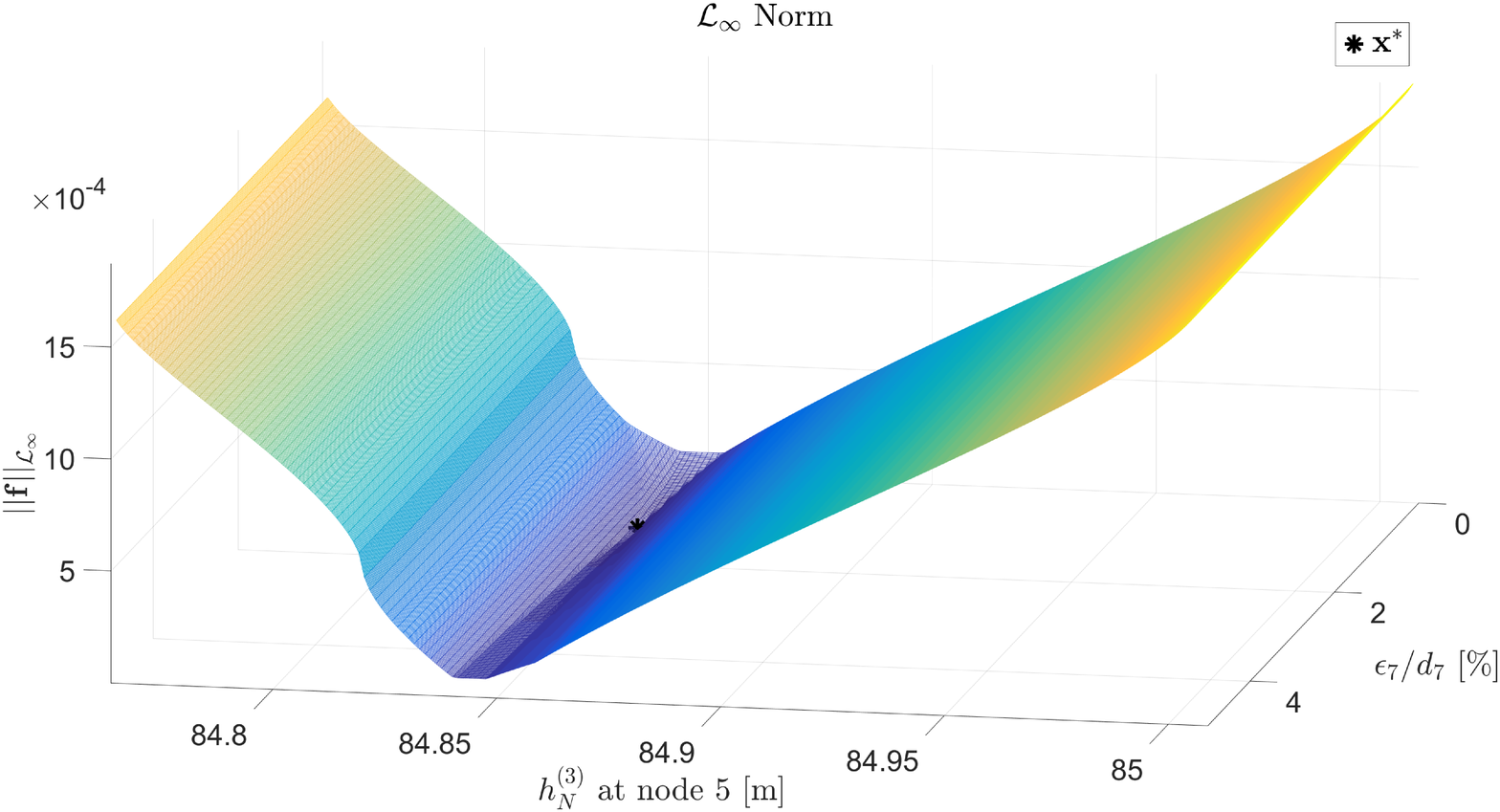}
            \caption{{\small $\mathcal{L}_{\infty}$ Norm.}}    
            \label{fig:SolSpace3}
        \end{subfigure}
\caption{Norm of $\bm{f}(\bm{x})$ plotted along the parameter space of $\epsilon_7/d_7$ and $h_{N,5}^{(3)}$ regarding the network in figure \ref{fig:3CycleNet}.}
 \label{fig:SolSpace}
\end{figure}

There is a particularly weak slope towards $\bm{x}^*$  along the $\epsilon_7/d_7$ axes in the $\mathcal{L}_2$ norm of figure \ref{fig:SolSpace1}, whereas the $\mathcal{L}_1$ norm in figure \ref{fig:SolSpace2} shows the overall highest slope towards the real root $\bm{x}^*$. This result is consistent, even when varying different roughnesses (one of the $n_{\uell}$ pipe roughnesses) and different $h^{(i)}_{N,j}$ in the variable space\label{2nd_observation}.

\subsection{Measurement-Sets}

The non-zero components of the nodal consumption are denoted by $\bm{q} = [\begin{matrix} q_2 &q_3 &q_4\end{matrix}]^T$ and can be obtained by $\bm{q} = \bm{C}_h \bar{\bm{q}}$ in this example.

\vspace{0.2cm}
\begin{figure}[H]
 \hspace{-0.2cm}
        \begin{subfigure}[htb]{0.475\textwidth}  
            \centering 
            \includegraphics[width=1.175\linewidth]{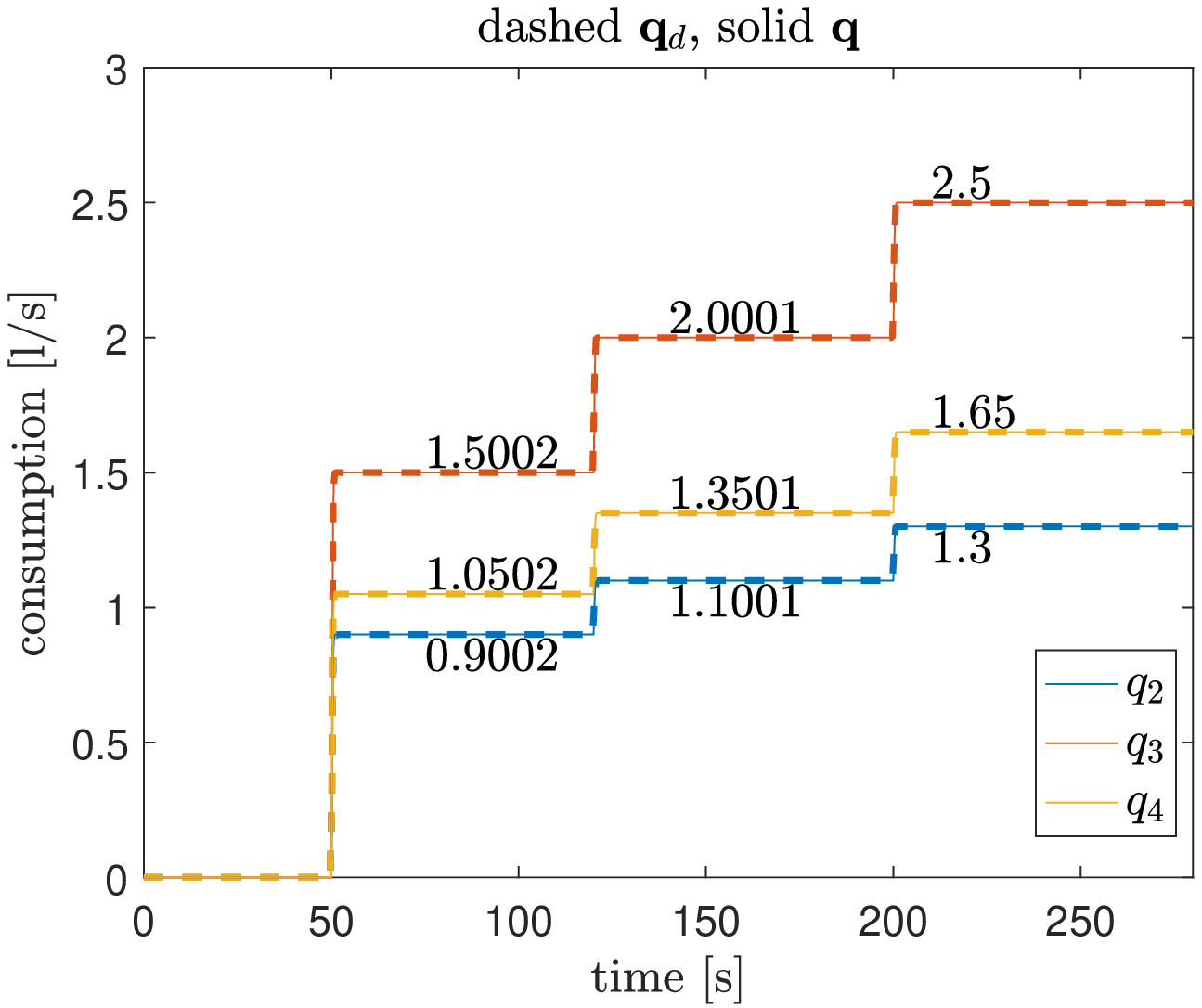}
            \caption{{\small Comparison of $\bm{q}_d(t)$ and $\bm{q}(t)$.}}    
            \label{fig:3cyclePlain_q}
        \end{subfigure}
        \quad
        \begin{subfigure}[htb]{0.475\textwidth}   
            \centering 
            \includegraphics[width=1.175\linewidth]{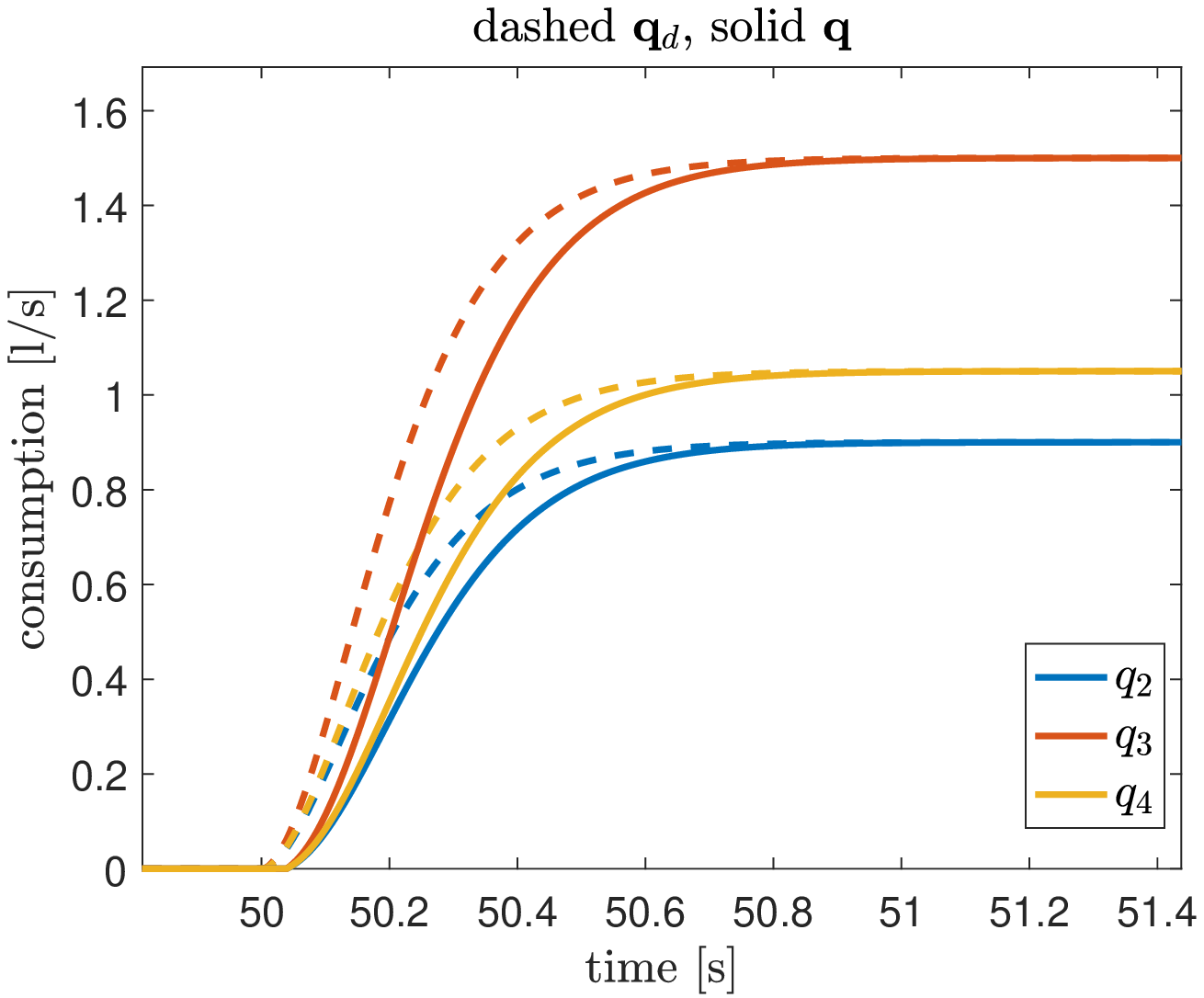}
            \caption{{\small Zoom of figure \ref{fig:3cyclePlain_q} showing dynamics.}}    
            \label{fig:3cyclePlain_qzoom}
        \end{subfigure}
\end{figure}
\begin{figure}[H]\ContinuedFloat    
        \begin{subfigure}[t]{0.475\textwidth}               
        \centering 
            \includegraphics[width=1.175\linewidth]{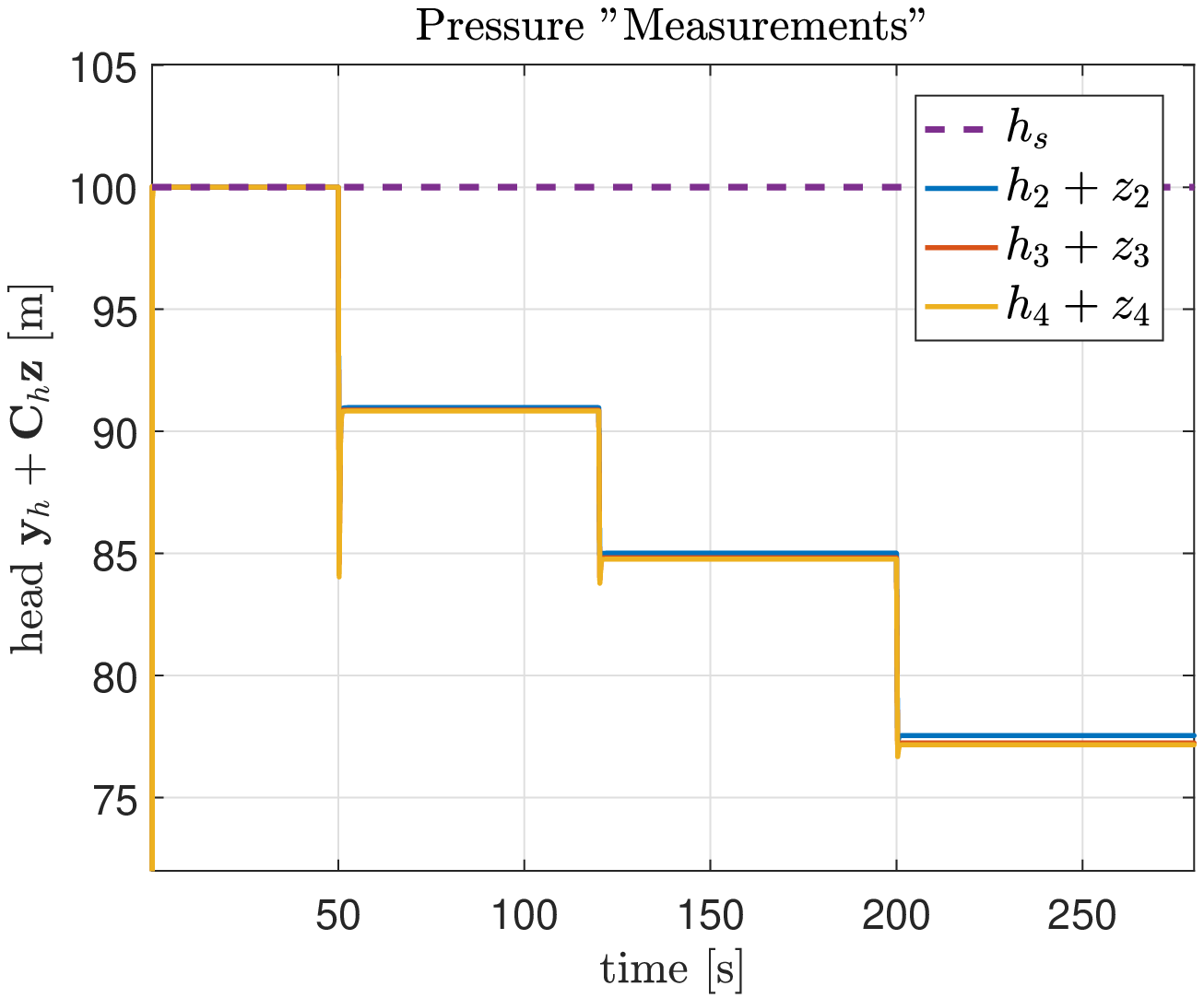}
            \caption{{\small Very little difference in $\bm{y}_h(t)+\bm{C}_h \bm{z}$}.}    
            \label{fig:3cyclePlain_yh}
        \end{subfigure}      
        \hspace{0.2cm}   
        \begin{subfigure}[t]{0.475\textwidth}   
            \centering 
            \includegraphics[width=1.175\linewidth]{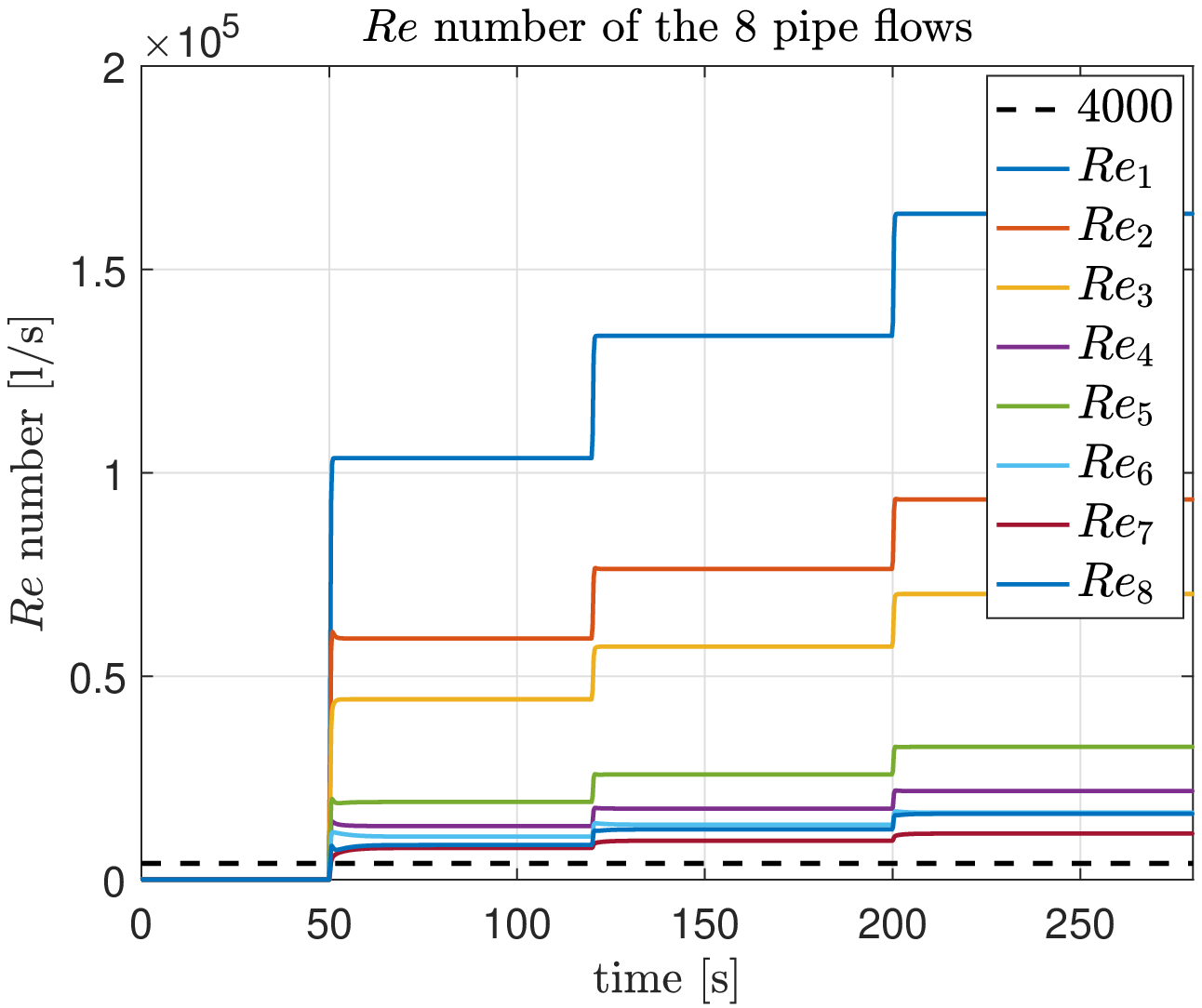}
            \caption{{\small Reynolds numbers.}}    
            \label{fig:3cyclePlain_Re}
        \end{subfigure}
\caption{Figures showing $n_{{\rm{m}}}=n_{{\rm{m,min}}} = 3$ ``measurement'' -sets for the roughness identification of the 3-cycle network in figure \ref{fig:3CycleNet}.}
\label{fig:3cyclePlain}
\end{figure}

Figure \ref{fig:3cyclePlain_q} shows that the real consumption $\bm{q}$ follows the reference $\bm{q}_d$, whereas the steady-state consumption values applied for \eqref{eq:calibrationTur} were displayed explicitly. As a remark, all the selected references concerning $\bm{q}_d(t)$ were generated with a filtered step using a \textit{Butterworth} filter which degree 2 and a cut-off angular frequency of $\omega_c = 10$ rad/s each concerning figure \ref{fig:3cyclePlain_q}.

 Beforehand, the selected configuration within consumption $\bm{q}(t)$, which leads to sensed head values $\bm{y}_h+ \bm{C}_h \bm{z}$, which can barely be distinguished among each other, was chosen on purpose for this example. Due to very little difference among the $\bm{y}_h + \bm{C}_h \bm{z}$, numerical inaccuracies are sufficient to cause serious difficulty to restore the roughness $\bm{\epsilon}$ with $\bm{y}_h$ and $\bm{q}$ when applying \eqref{eq:calibrationTur}, presumably violating Assumption \ref{ass:linearIndependence}. In this context it is important to emphasize that this illustrative example was configured such that all flows in all the 3 ``measurement''-sets are in the turbulent regime according to Assumption \ref{ass:FullTurbulent}, seen in figure \ref{fig:3cyclePlain_Re}. The quantities to set up \eqref{eq:calibrationTur} as well as its \textit{Jacobian} \eqref{eq:Jacobian} are summarized in the following table.

\vspace{0.3cm}
\begin{table}[H]                                                                       
\centering                                                                                
\begin{tabular}{c|c|c|c||c}   
set &1 &2 &3  &unit                   \\                             
\hline             \hline  
   &90.9743   &85.0087   &77.5380&\\
$\bm{y}_h$   &90.8720   &84.8200   &77.2370& m\\
   &90.8339   &84.7638   &77.1594 &\\
   \hline
    &0.9002    &1.1001    &1.3000&\\
 $\bm{q}$   &1.5002    &2.0001    &2.5000& l/s\\
    &1.0502    &1.3501    &1.6500& \\  
    \hline
$\bm{h}_s$    &100 &100 &100&m                                         
\end{tabular}                                                                             
\caption{Measurement-sets.}  
\label{tab:3cycle_config1}                                                                
\end{table}
%

\paragraph{Initial Values}

The initial value for the not-measured pressure head at node $5$, i.e. $h_{N_0,5}^{(i)}$, is chosen as the mean over all surrounding pressure heads (which happen to be located at nodes with pressure sensors).
\begin{subequations}
\label{eq:x0_3cycle_hNrange}
\begin{gather}
h_{N_0,5}^{(i)} = \frac{1}{n_{{\rm{j}}}-n_{{\rm{p}}}}\sum_{j=1}^{n_{{\rm{j}}}-n_{{\rm{p}}}} [\bm{y}_h]_j^{(i)} = \frac{1}{3}\sum_{j=2}^{4} h_j^{(i)} \qquad \forall i
\end{gather}
The initial value for the not-measured pressure head at node $1$, i.e. $h_{N_0,1}^{(i)}$, is chosen analogously such that
\begin{gather}
h_{N_0,1}^{(i)} = \frac{1}{3}\left( h_s^{(i)} + h_2^{(i)}+h_3^{(i)} \right) \qquad \forall i .
\end{gather}
\end{subequations}
The initial roughness value is chosen as 1\% of the pipes' diameter, leading to the initial vector
\begin{align}
\bm{x}_0 &= \left[
\begin{matrix}
 \bm{\epsilon}_0^T & h_{N_0,1}^{(1)}& h_{N_0,5}^{(1)} & h_{N_0,1}^{(2)}&  h_{N_0,5}^{(2)} &h_{N_0,1}^{(3)} & h_{N_0,5}^{(3)}
\end{matrix}
\right]^T \\
&=
\left[
\begin{matrix}
0.0004  \times \bm{1}^T_{n_{\uell}} 
  &93.9488
  &90.8934
   &89.9429
   &84.8642
   &84.9250
   &77.3115
\end{matrix}
\right]^T \nonumber
\label{eq:x0_3cycle_calibrate}
\end{align}
for launching Algorithm \ref{alg:IC}. The minimal and maximal value of all surrounding pressure heads  in the corresponding measurement-set is chosen for lower and upper boundary concerning $\underline{\bm{h}_N}$ and $\overline{\bm{h}_N}$ , leading, for instance, to a maximal value of the pressure at node 1 of $h^{(i)}_{N,1} \le \overline{h_{N}}_{1}^{(i)} = h_s^{(i)} = 100 \, \forall i$. As it will turn out that the presented $\bm{x}_{h_N}$ results never leave their defined physically relevant range, these boundaries are not important for the present example. 

\subsection{Results and Discussion}

The initial values along some iterations of Algorithm \ref{alg:IC} are presented in table \ref{tab:3cycle_x0} whereas
table \ref{tab:3cycle_x} presents the solutions of Algorithm \ref{alg:Newton}, which was launched with initial values in table  \ref{tab:3cycle_x0}, along some iterations of Algorithm \ref{alg:IC}. Concerning Algorithm \ref{alg:Newton}, the maximal number of iterations was limited to $1000$, whereas $\epsilon_f = 10^{-7}$ and $\epsilon_x = 5 \times 10^{-7}$.  

\vspace{0.2cm}
%
\begin{table}[H]                                                                       
\hspace{-0.6cm}                                                                                
\begin{tabular}{c||c|c|c|c|c|c|c||c}   
iteration &1 &2 &3 &4 &5 &6 &7 &$\bm{x}^*$                      \\                             
\hline             \hline                                                                       
$\epsilon_{1}$ & 0.400 & 0.225 & 1.272 & 1.272 & 0.319 & 1.177 & 1.239 & 2.000 \\         
\hline                                                                                    
$\epsilon_{2}$ & 0.400 & 1.632 & 1.562 & 1.567 & 1.730 & 1.730 & 1.730 & 1.750 \\         
\hline                                                                                    
$\epsilon_{3}$ & 0.400 & 1.410 & 1.227 & 1.083 & 1.496 & 1.496 & 1.496 & 1.500 \\         
\hline                                                                                    
$\epsilon_{4}$ & 0.400 & 1.111 & 1.361 & 1.361 & 1.176 & 1.176 & 1.176 & 1.250 \\         
\hline                                                                                    
$\epsilon_{5}$ & 0.400 & 1.054 & 0.824 & 0.824 & 1.028 & 1.028 & 1.028 & 1.000 \\         
\hline                                                                                    
$\epsilon_{6}$ & 0.400 & 0.788 & 0.696 & 0.696 & 0.786 & 0.786 & 0.786 & 0.750 \\         
\hline                                                                                    
$\epsilon_{7}$ & 0.400 & 0.499 & 0.471 & 0.471 & 0.505 & 0.505 & 0.505 & 0.500 \\         
\hline                                                                                    
$\epsilon_{8}$ & 0.400 & 0.250 & 0.228 & 0.228 & 0.250 & 0.250 & 0.250 & 0.250 \\         
\hline                                                                                    
$h_{N,1}^{(1)}$ & 93.949 & 93.047 & 94.344 & 94.344 & 93.097 & 93.097 & 93.097 & 93.104 \\
\hline                                                                                    
$h_{N,5}^{(1)}$ & 90.893 & 90.885 & 90.886 & 90.886 & 90.885 & 90.885 & 90.885 & 90.885 \\
\hline                                                                                    
$h_{N,1}^{(2)}$ & 89.943 & 88.443 & 90.602 & 90.602 & 88.525 & 88.525 & 88.525 & 88.538 \\
\hline                                                                                    
$h_{N,5}^{(2)}$ & 84.864 & 84.846 & 84.848 & 84.848 & 84.846 & 84.846 & 84.846 & 84.846 \\
\hline                                                                                    
$h_{N,1}^{(3)}$ & 84.925 & 82.674 & 85.916 & 85.916 & 82.799 & 82.799 & 82.799 & 82.818 \\
\hline                                                                                    
$h_{N,5}^{(3)}$ & 77.311 & 77.280 & 77.283 & 77.283 & 77.280 & 77.280 & 77.280 & 77.280 \\                                                              
\end{tabular}                                                                             
\caption{Initial values for calibrating the 3-cycle network (figure \ref{fig:3CycleNet}) via Algorithm \ref{alg:Newton} along iterations of Algorithm \ref{alg:IC}. Roughnesses $\epsilon_i$ are presented in mm, whereas pressure heads are presented in m.}  
\label{tab:3cycle_x0}                                                                
\end{table} 
\vspace{0.2cm}

\begin{table}[H]                                                                       
\hspace{-0.6cm}                                                                                
\begin{tabular}{c||c|c|c|c|c|c|c||c}     
iteration &1 &2 &3 &4 &5 &6 &7 &$\bm{x}^*$                      \\           
\hline \hline                                                                
$\epsilon_{1}$ & \cellcolor{blue!15}2.036 & 1.272 & 1.986 &  \cellcolor{blue!15}2.005 & 1.340 &  \cellcolor{blue!15}2.031 &  \cellcolor{blue!15}2.036 & 2.000 \\              
\hline                                                                                         
$\epsilon_{2}$ & 1.632 &  \cellcolor{blue!15}4.280 & 1.802 & 1.730 &  \cellcolor{blue!15}4.188 & 1.648 & 1.631 & 1.750 \\              
\hline                                                                                         
$\epsilon_{3}$ & 1.410 & \cellcolor{blue!15} 3.880 & 1.562 & 1.496 &  \cellcolor{blue!15}3.398 & 1.425 & 1.408 & 1.500 \\              
\hline                                                                                         
$\epsilon_{4}$ & 1.111 & 1.361 & 1.216 & 1.176 &  \cellcolor{blue!15}3.568 & 1.183 & 1.185 & 1.250 \\              
\hline                                                                                         
$\epsilon_{5}$ & 1.054 & 0.824 & 0.990 & 1.028 & 0.393 & 1.019 & 1.019 & 1.000 \\              
\hline                                                                                         
$\epsilon_{6}$ & 0.788 & 0.696 & 0.752 & 0.786 & 0.199 & 0.755 & 0.756 & 0.750 \\              
\hline                                                                                         
$\epsilon_{7}$ & 0.499 & 0.471 & 0.495 & 0.505 & 0.389 & 0.493 & 0.494 & 0.500 \\              
\hline                                                                                         
$\epsilon_{8}$ & 0.250 & 0.228 & 0.261 & 0.250 & 0.710 & 0.265 & 0.264 & 0.250 \\              
\hline                                                                                         
$h_{N,1}^{(1)}$ & 93.047 & 94.344 & 93.128 & 93.097 & 94.219 & 93.056 & 93.047 & 93.104 \\     
\hline                                                                                         
$h_{N,5}^{(1)}$ & 90.885 & 90.886 & 90.885 & 90.885 & 90.890 & 90.885 & 90.885 & 90.885 \\     
\hline                                                                                         
$h_{N,1}^{(2)}$ & 88.443 & 90.602 & 88.578 & 88.525 & 90.395 & 88.458 & 88.442 & 88.538 \\     
\hline                                                                                         
$h_{N,5}^{(2)}$ & 84.846 & 84.848 & 84.846 & 84.846 & 84.856 & 84.846 & 84.846 & 84.846 \\     
\hline                                                                                         
$h_{N,1}^{(3)}$ & 82.674 & 85.916 & 82.877 & 82.799 & 85.604 & 82.697 & 82.674 & 82.818 \\     
\hline                                                                                         
$h_{N,5}^{(3)}$ & 77.280 & 77.283 & 77.281 & 77.280 & 77.297 & 77.280 & 77.280 & 77.280 \\     
\hline         \hline                                                                               
$v(\bm{x}) \times 10^{5}$ & 5.897 & 1.969 & 7.630 & \cellcolor{olive!20}1.932 & 5014.8 & 5.151 & 5.089 & 0.011 \\
\hline                                                                                    
\end{tabular}                                                                             
\caption{Solutions of \eqref{eq:calibrationTur} concerning the 3-cycle network (figure \ref{fig:3CycleNet}) by Algorithm \ref{alg:Newton} along iterations of Algorithm \ref{alg:IC}. The corresponding initial values can be found in table \ref{tab:3cycle_x0}. Roughnesses $\epsilon_i$ are presented in mm, whereas pressure heads are presented in m.}                                                                  
\label{tab:3cycle_x}                                                                
\end{table}  
                    
In reference to table \ref{tab:3cycle_x0} and \ref{tab:3cycle_x}, Algorithm \ref{alg:IC} ran for a fixed number of iterations, namely 7 iterations in order allow visible investigation of its working principle. In this context, the accuracy limits $\epsilon_f$ and $\epsilon_x$ for the while loop in line \ref{alg:IC_whileCondition} of Algorithm \ref{alg:IC} were chosen such that Algorithm \ref{alg:IC} does not abort until the fixed iteration-number 7 was reached.           \\\


All roughnesses and the not-measured pressure heads in all measurement-sets could be restored with reasonable accuracy (i.e. a maximal deviation of 6\% concerning $\epsilon_4^{+}$). The intermediate best result $\bm{x}^+$, in reference to Algorithm \ref{alg:IC}, in table \ref{tab:3cycle_x} is colored in (color) olive with a residual of $v(\bm{x}^+) = 1.932 \times10^{-5}$ m$^3$/s $= 1.932 \times 10^{-2}$ l/s, whereas the real root can still clearly be distinguished from all other solutions of Algorithm \ref{alg:Newton} with a residual of $v(\bm{x}^*) = 0.011 \times 10^{-2}$ l/s. The blue colored values in table \ref{tab:3cycle_x} are those roughnesses which exceed the 5\% mark of the corresponding pipe's diameter. Those roughnesses are then selected by a random number generator applied in Algorithm \ref{alg:IC} for the next iteration. Randomly generated roughnesses can be found in corresponding entries of table \ref{tab:3cycle_x0}.

%
On the contrary, one has to pay attention to the fact that a solution was found which features an ever so slightly higher residual $v(\bm{x}) = 1.969 \times 10^{-2}$ l/s in the second iteration of table \ref{tab:3cycle_x} compared to $v(\bm{x}^+)= 1.932 \times 10^{-2}$ l/s.
When also considering measurement noise, one certainly loses the capability to identify the real root by only looking at $v(\bm{x})$. Also, the indicator that the solution in the second iteration has two roughnesses $\epsilon_2,\epsilon_3$ which exceed their physical bounds in comparison to $\bm{x}^+$ (olive), only featuring $\epsilon_1$ which exceeds $0.05 d_1$ by a mere $0.25$\%, will not be sufficient. In the opinion of the authors, the only possibility to deal with measurement noise and potentially non-zero minor losses (referring to Assumption \ref{ass:NoiseCalibration} and \ref{ass:minorLossesCalibration}) is by considering measurement-sets which are independent from each other in a sense that ``measured'' heads $\bm{y}^{(i)}_h + \bm{C}_h \bm{z}$ are sufficiently different from each other. This comes on top of the requirement formulated within Assumption \ref{ass:NoiseCalibration}. In this context, it turned out particularly useful to not only consider the minimal number of required measurement-sets $n_{{\rm{m,min}}}$ but additional, thereby improving the number of (nodal) equations to the number of unknowns. Nevertheless, the solving becomes more delicate due to a fast growing equation-set \eqref{eq:calibrationTur}.

\section{Conclusion and Outlook}

This manuscript focused on the deduction of circumstances which allow individual pipe roughness parameters to be \textit{uniquely} reconstructed from the commonly applied sensor configuration. It turned out that a set of independent measurements is needed to accommodate for the large number of unknowns. However, the proposed algorithms enable to find the real root of the equation-set reliably, provided that independent measurement-sets are available.

Before this methodology can be applied to real-world networks, the formulation has to be extended to also allow pipe flows in the laminar and transitional \textit{Reynolds} area. Therefore, a sufficiently smooth and explicit description of the flow in the transitional \textit{Reynolds} as in \cite{transitionalWaterFlow} is required. which satisfies not only the boundary conditions to \textit{Colebrook \& White}'s flow \eqref{eq:ft}, but also the gradient with respect to the roughness and the head loss.




%
%
%
%
%



\bibliography{example}

\end{document}